\documentclass[11pt]{article}
\input psfig.sty

\usepackage{graphics}
\usepackage{cite} % collapse adjacent citations

\psdraft
\psfigurepath{/home/httpd/html/mrenna/lc/}

\newenvironment{Eqnarray}%
     {\arraycolsep 0.14em\begin{eqnarray}}{\end{eqnarray}}

\makeatletter
\@addtoreset{equation}{section}

\makeatother

\def\NPB#1#2#3{{Nucl.~Phys.} B {\bf{#1}}, #3 (19#2)}
\def\PLB#1#2#3{{Phys.~Lett.} B {\bf{#1}}, #3 (19#2)}
\def\PRD#1#2#3{{Phys.~Rev.} D {\bf{#1}}, #3 (19#2)}
\def\PRL#1#2#3{{Phys.~Rev.~Lett.} {\bf{#1}}, #3 (19#2)}

\newcommand{\be}{\begin{equation}}
\newcommand{\ee}{\end{equation}}
\newcommand{\bea}{\begin{Eqnarray}}
\newcommand{\eea}{\end{Eqnarray}}

\def\lsim{\mathrel{\raise.3ex\hbox{$<$\kern-.75em\lower1ex\hbox{$\sim$}}}}
\def\gsim{\mathrel{\raise.3ex\hbox{$>$\kern-.75em\lower1ex\hbox{$\sim$}}}}
\def\ifmath#1{\relax\ifmmode #1\else $#1$\fi}
\def\ls#1{\ifmath{_{\lower1.5pt\hbox{$\scriptstyle #1$}}}}
\def\missET{\not\!\!E_T}
\def\misspT{\not\!\!p_T}

\begin{document}

\pagestyle{empty}
\begin{flushright}
ANL-HEP-PR-02-042  \\
EFI-02-88 \\
FERMILAB-Pub-02/072-T  \\
hep-ph/0206167 \\
\end{flushright}

\vskip1cm

\renewcommand{\thefootnote}{\fnsymbol{footnote}}
\begin{center}
{\Large\bf 
Top-squark searches at the Tevatron in models
of low-energy supersymmetry breaking
}\\[1cm]
{\large Marcela Carena$^a$, Debajyoti Choudhury$^b$, Rodolfo A. Diaz$^c$,\\
Heather E. Logan$^a$ and Carlos E.M. Wagner$^{d,e}$~\footnote{Electronic
addresses:
carena@fnal.gov, debchou@mri.ernet.in, radiaz@ciencias.unal.edu.co,
logan@fnal.gov,\\ cwagner@hep.anl.gov}
}\\[6pt]
{\it
$^a$ Theoretical Physics Department, Fermilab, PO Box 500, 
Batavia, Illinois 60510-0500, USA\\
$^b$ Harish-Chandra Research Institute, Chhatnag Road, Jhusi,
Allahabad 211 019, India\\
$^c$ Departamento de Fisica, Universidad Nacional de Colombia,
Bogota, Colombia\\
$^d$ HEP Division, Argonne National Lab, 9700 S. Cass Ave,
Argonne, Illinois 60439, USA\\
$^e$ Enrico Fermi Institute, University of Chicago, 5640 Ellis Ave, 
Chicago, Illinois 60637, USA}\\[.5cm]

{\bf Abstract}
\end{center}

\noindent

We study the production and decays of top squarks (stops) at the Tevatron 
collider in models of low-energy supersymmetry breaking. We consider
the case where the lightest Standard Model (SM) superpartner is a light
neutralino that predominantly decays into a photon and a light gravitino.
Considering the lighter stop to be the next-to-lightest Standard Model
superpartner, we analyze stop signatures associated with jets, photons
and missing energy, which lead to signals naturally larger than the
associated SM backgrounds. We consider both 2-body and 3-body decays of 
the top squarks and show that the reach of the Tevatron can be
significantly larger than that expected within either the 
standard supergravity models or models of low-energy supersymmetry breaking
in which the stop is the lightest SM superpartner.
For a modest
projection of the final Tevatron luminosity, ${\cal{L}} \simeq 4$ 
fb$^{-1}$, stop masses of order
300 GeV are accessible at the Tevatron collider in both
2-body and 3-body decay modes.
We also consider the production and decay of ten degenerate squarks
that are the supersymmetric partners of the five light quarks.
In this case we find that common squark masses up to 360 GeV are easily
accessible at the Tevatron collider, and that the reach increases
further if the gluino is light.\\[10pt]

\vfill
\clearpage
%%%%%%%%%%%%%%%%%%%%%%%%%%%%%%%%%%%%%%%%%%%%%%%%%%%%%%%%%%%%%%%%%%%%%%%%

\renewcommand{\thefootnote}{\arabic{footnote}}
\setcounter{footnote}{0}

%%%%%%%%%%%%%%%%%%%%%%%%%%%%%%%%%%%%%%%%%%%%%%%%%%%%%%%%%%%%%%%%%%%%%%%%%%%%
\pagestyle{plain}

\section{Introduction}

The Standard Model with a light Higgs boson provides a very good
description of all experimental data.  The consistency of the precision
electroweak data with the predictions of the Standard Model suggests that,
if new physics is present at the weak scale, it is most probably 
weakly interacting and consistent with the presence of a light Higgs
boson in the spectrum. Extensions of the Standard Model based on softly
broken low energy supersymmetry (SUSY) \cite{HaberKaneNilles}
provide the most attractive
scenarios of physics beyond the Standard Model fulfilling these properties.
If the supersymmetry breaking masses are of the order of 1 TeV,
supersymmetry stabilizes the hierarchy between the 
Planck scale $M_P$ and the electroweak scale. 
Furthermore, the minimal supersymmetric extension
of the Standard Model (MSSM) significantly improves the precision with
which the three gauge couplings unify and leads to the presence of
a light Higgs boson with a mass below 135 GeV \cite{lightHiggsmass}.

Perhaps the most intriguing property of supersymmetry is that
local supersymmetry naturally leads to the presence of gravity (supergravity).
In the case of global supersymmetry,
a massless spin one-half particle, the Goldstino, appears
in the spectrum when supersymmetry is spontaneously broken. 
In the case of local supersymmetry, the Goldstino provides the additional
degrees of freedom to make the gravitino a massive particle~\cite{Fayet}. 
In the simplest scenarios,
the gravitino mass $m_{\tilde G}$ is directly proportional
to the square of the supersymmetry breaking scale $\sqrt{F_{\rm SUSY}}$:
\begin{equation}
        m_{\tilde{G}} \simeq F_{\rm SUSY}/\sqrt{3} M_P.
\label{eq:mgrav}
\end{equation}

The relation between the supersymmetry breaking scale $\sqrt{F_{\rm SUSY}}$
and the supersymmetry particle masses depends on the specific supersymmetry
breaking mechanism.  In general, the superpartner masses $M_{\rm SUSY}$ 
are directly proportional to $F_{\rm SUSY}$ and inversely
proportional to the messenger scale $M_m$ at which the supersymmetry
breaking is communicated to the visible sector:
\begin{equation}
        M_{\rm SUSY} \simeq \alpha \frac{F_{\rm SUSY}}{M_m},
\label{eq:Msusy}
\end{equation}
where $\alpha$ is the characteristic coupling constant coupling the 
messenger sector to the visible one. If the breakdown of supersymmetry
is related to gravity effects,
$M_m$ is naturally of the order of the Planck scale and $\alpha$ is
of order one, 
so that $M_{\rm SUSY}$ is at the TeV scale if $\sqrt{F_{\rm SUSY}} 
\sim 10^{11}$ GeV.  
In gauge mediated scenarios (GMSB)~\cite{gms,DineLESB}, 
instead, the couplings $\alpha$ are associated with the Standard Model
gauge couplings times a loop suppression factor, so that  
values of $F_{\rm SUSY}/M_m$ of order 100 TeV yield masses
for the lighter Standard Model superpartners of order 100 GeV.

When relevant at low energies, the gravitino interactions with matter
are well described through the interactions 
of its spin 1/2 Goldstino component~\cite{Fayet}.
The Goldstino has derivative couplings with the visible sector
with a strength proportional to $1/F_{\rm SUSY}$. 
In scenarios with a high messenger scale, of order $M_P$, 
Eqs.~\ref{eq:mgrav} and \ref{eq:Msusy} imply that the gravitino has a mass
of the same order as the other SUSY particles,
and its interactions are extremely weak.  In such scenarios, the
gravitino plays no role in the low-energy phenomenology.
However, in low energy supersymmetry breaking scenarios such as GMSB
in which the messenger scale is significantly lower than the 
Planck scale, the supersymmetry breaking scale is much smaller.
Typical values in the GMSB case are $M_m \sim 10^5 - 10^8$ GeV,
leading to a supersymmetry breaking scale $\sqrt{F_{\rm SUSY}}$ 
roughly between $10^5$ and a few times $10^6$ GeV.
The gravitino then becomes significantly lighter than the superpartners
of the quarks, leptons and gauge bosons,
and its interaction strength is larger.
As the lightest supersymmetric particle (LSP), the gravitino must 
ultimately be produced at the end of all superparticle decay chains
if $R$-parity is conserved 
(for an analysis of the case of $R$-parity violation see Ref.~\cite{rpgm}).

Depending on the strength of the gravitino coupling, the decay length of
the next-to-lightest supersymmetric particle (NLSP) can be large 
(so that the NLSP is effectively stable from the point of view of 
collider phenomenology; this occurs when 
$\sqrt{F_{\rm SUSY}} \gsim 1000$ TeV), intermediate (so that
the NLSP decays within the detector giving rise to spectacular
displaced vertex signals; this occurs when 
1000 TeV $\gsim \sqrt{F_{\rm SUSY}} \gsim 100$ TeV), or microscopic 
(so that the NLSP decays promptly; this occurs when 
$\sqrt{F_{\rm SUSY}} \lsim 100$ TeV)~\cite{SUSYRun2,at2,KaneTev}.
The decay branching fractions of the Standard Model
superpartners other than the NLSP into the gravitino are typically 
negligible.
However, if the supersymmetry breaking scale is very low,
$\sqrt{F_{\rm SUSY}} \ll 100$ TeV (corresponding to a gravitino mass
$\ll 1$ eV), then the gravitino coupling strength
can become large enough that superpartners other than the NLSP may 
decay directly into final states containing a gravitino \cite{KaneTev,Zwirner}.
In any case, the SUSY-breaking scale must be larger than the mass of 
the heaviest superparticle; an approximate lower bound of 
$\sqrt{F_{\rm SUSY}} \simeq 1$ TeV yields a gravitino mass of about 
$10^{-3}$ eV.

In many models, the lightest supersymmetric partner of a Standard Model
particle is a neutralino, $\tilde \chi^0_1$. 
The decay widths of the $\tilde \chi^0_1$
into various final state particles are given by:
\begin{equation}
        \Gamma(\tilde{\chi}_1^0 \to X \tilde{G})   
        \simeq K_X N_X \frac{m_{\tilde \chi_1^0}}{96 \pi} 
        \left(\frac{m_{\tilde \chi_1^0}}
        {\sqrt{M_P m_{\tilde{G}}}}\right)^4 
        \left( 1 - \frac{m_X^2}{m_{\tilde \chi_1^0}^2} \right)^4, 
\label{rategm}
\end{equation}
where $K_X$ is a projection
factor equal to the  square of the component in the NLSP of the superpartner
of the particle $X$, and $N_X$ is the number of degrees of freedom of $X$.
If $X$ is a photon, for instance, $N_X = 2$ and
\begin{equation}
        K_X = \left| N_{11} \cos\theta_W + N_{12} \sin\theta_W \right|^2,
\end{equation}
where $N_{ij}$ is the mixing matrix connecting the 
neutralino mass eigenstates
to the weak eigenstates in the basis $\tilde{B}, \tilde{W}, \tilde{H}_1,
\tilde{H}_2$. 

If the neutralino has a significant photino component, it will lead to 
observable decays into photon and gravitino. 
Since the heavier supersymmetric particles decay into the NLSP, which
subsequently decays into photon and gravitino, supersymmetric
particle production will be characterized
by events containing photons and missing energy.
This is in contrast to 
supergravity scenarios, where, unless very specific conditions are
fulfilled~\cite{radneu,m2eqm1}, 
photons do not represent a characteristic signature.
The presence of two energetic photons plus 
missing transverse energy provides a distinctive SUSY signature with 
very little Standard Model background.

If $\tilde \chi^0_1$ decays with a large branching ratio
into photons, its mass is severely constrained by LEP data.
However, this bound is model-dependent: there is no tree-level coupling 
of the $\gamma$ and $Z$ to two binos or to two neutral winos, but
LEP could produce bino or neutral wino pairs through $t$-channel exchange of 
selectrons. 
Since the bino is associated with the smallest of all gauge interactions 
and, in addition, its mass is more strongly renormalized downward at 
smaller scales compared to the wino mass, 
in many models the lightest neutralino has a significant
bino component.  
Therefore, if the NLSP is approximately a pure bino,
the bounds on its mass depend strongly on the selectron mass. 
For selectron masses below 200 GeV, the present bound on such 
a neutralino is of about 90 GeV~\cite{LEPSUSYWG}. 
Then, as emphasized before, it will decay
via $\tilde \chi^0_1 \to \gamma \tilde G$.
If $\tilde \chi^0_1$ is heavy enough, then the decays 
$\tilde \chi^0_1 \to Z \tilde G$ and $\tilde \chi^0_1 \to h^0 \tilde G$
are also kinematically allowed; however, the decay widths into 
these final states are kinematically suppressed compared
to the $\gamma \tilde G$ final state and will only be important
if the photino component of $\tilde \chi^0_1$ is small or 
if $\tilde \chi^0_1$ is significantly heavier than $Z$ and 
$h^0$ \cite{KaneTev}. 

In most SUSY models it is natural for the lighter top squark, $\tilde t_1$,
to be light compared to the other squarks.
In general, due to the large top Yukawa coupling there is a large 
mixing between the weak eigenstates $\tilde t_L$ and $\tilde t_R$,
which leads to a large splitting between the two stop mass eigenstates.  
In addition, even if all squarks have a common mass at the messenger scale, 
the large top Yukawa coupling typically results in 
the stop masses being driven (under renormalization group evolution)
to smaller values at the weak scale.
An extra motivation to consider light third generation squarks comes
from the fact that 
light stops, with masses of about or smaller than the top quark mass, 
are  demanded for the realization of the mechanism
of electroweak baryogenesis within the context of the MSSM \cite{EWBG}.

In this paper, we examine in detail the production and decay of top squarks 
at Run II of the Tevatron collider in low-energy SUSY breaking 
scenarios, by assuming that the NLSP is the lightest neutralino,
which decays promptly to $\gamma \tilde G$.
We also investigate the production and decay of the other squarks,
and provide an estimate of the reach of Run II of the Tevatron in
the heavy gluino limit.
We work in the context of a general SUSY model
in which the SUSY particle masses
are {\it not} constrained by the relations predicted in the minimal
GMSB models.
We assume throughout that the gravitino coupling is strong enough
(or, equivalently, that the scale of SUSY breaking is low enough) 
that the NLSP decays promptly.
This implies an upper bound on the supersymmetry breaking scale
of a few tens to a few hundred TeV~\cite{at2,KaneTev,SUSYRun2},
depending on the mass of the NLSP.  
Our analysis can be extended to higher supersymmetry breaking 
scales for which the NLSP has a finite decay length, although in
this case some signal will be lost due to NLSP decays outside the
detector.  At least 50\% of the diphoton signal cross section remains 
though for NLSP decay lengths $c\tau \lsim 40$ cm~\cite{SUSYRun2}; this 
corresponds to a supersymmetry breaking scale below a few hundred
to about a thousand TeV, depending on the mass of the NLSP. 
Moreover, the displaced vertex associated with a finite decay length 
could be a very good additional discriminator for the signal. Thus, in
totality, our choice is certainly not an overly optimistic one. 

This paper is organized as follows.  In Sec.~\ref{sec:production} we
review the stop pair production cross section at the Tevatron. 
In Sec.~\ref{sec:literature} we summarize previous studies of stop
production and decay at Run II of the Tevatron.
In Sec.~\ref{sec:stopBRs} we discuss the SUSY parameter space
and the relative partial widths of the various stop decay modes.
In Sec.~\ref{sec:signals} we describe the signal for each of the
stop decay modes considered.  We describe the backgrounds
and the cuts used to separate signal from background in each case, 
and give signal cross sections after cuts.  This gives the reach
at Run~II.  We also note the possibility that stops can be produced
in the decays of top quarks.
In Sec.~\ref{sec:10squarks} we consider 
the production and decay of 10 degenerate 
squarks that are the supersymmetric partners of the five light quarks.
In Sec.~\ref{sec:conclusions} we summarize our conclusions.

%------------------------------------------------------------------
\section{Top squark production at Tevatron Run~II}
\label{sec:production}

Top squarks are produced at hadron colliders overwhelmingly via the strong
interaction, so that the tree-level cross sections are model independent
and depend only on the stop mass.
The production modes of the lighter stop, $\tilde t_1$, at the Tevatron are
$q \bar q \to \tilde t_1 \tilde t_1^*$ and $gg \to \tilde t_1 \tilde t_1^*$.
The cross sections for these processes are well known at leading
order (LO) \cite{Harrison,DEQ}, and the next-to-leading order (NLO) 
QCD and SUSY-QCD corrections have been computed \cite{PROSPINOstops}
and significantly reduce the renormalization scale dependence.
The NLO cross section is implemented numerically in
PROSPINO \cite{PROSPINO,PROSPINOstops}.

\begin{figure}
\resizebox{\textwidth}{!}{
\rotatebox{270}{\includegraphics{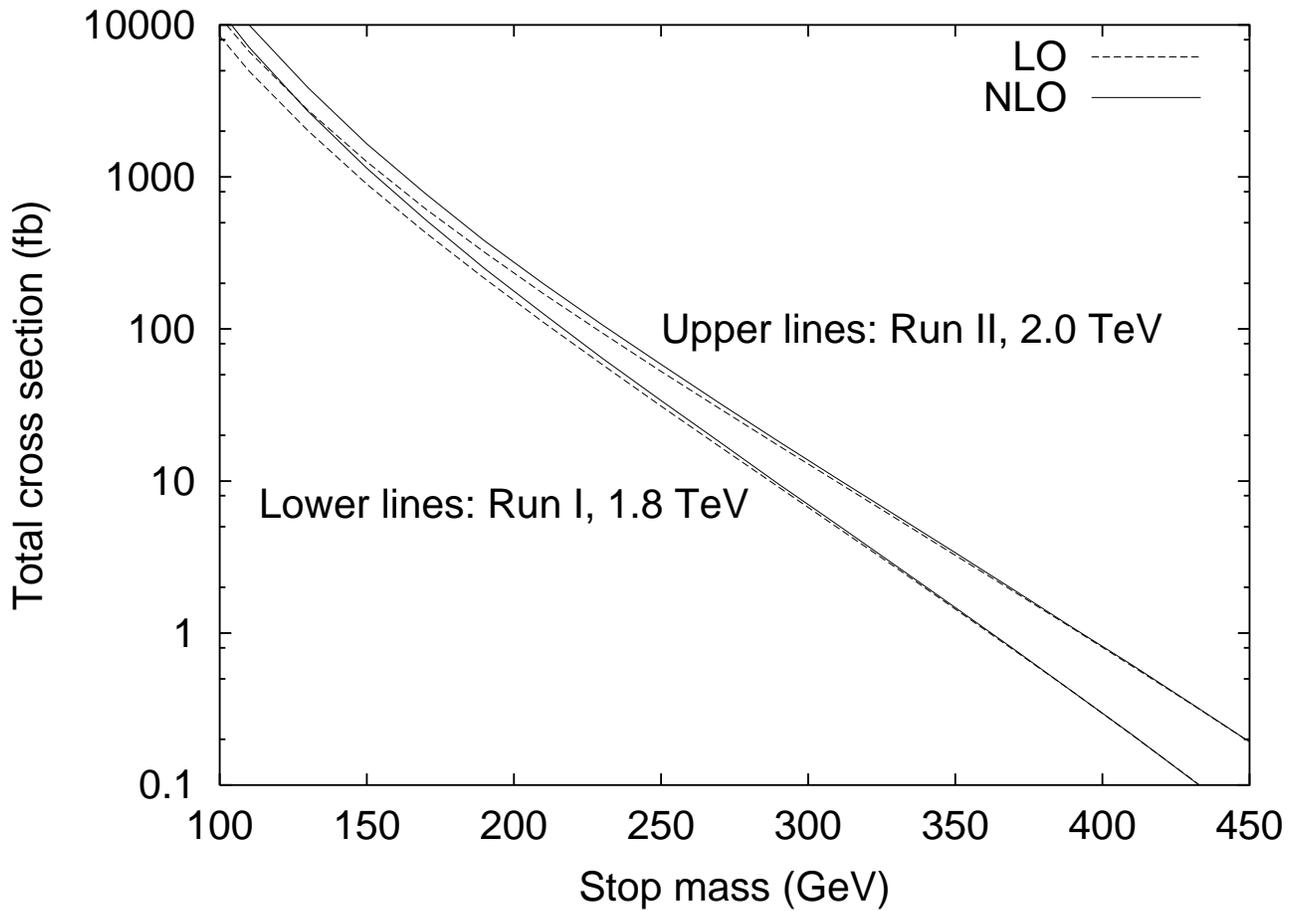}}}
\caption{LO and NLO cross sections
for stop pair production in $p \bar p$ collisions 
at Tevatron Run I (1.8 TeV) and Run II (2.0 TeV),
from PROSPINO~\cite{PROSPINO,PROSPINOstops}.  
Cross sections are evaluated at the scale $\mu = m_{\tilde t}$.
}
\label{fig:xsecmass}
\end{figure}

We generate stop events using the LO cross section evaluated at 
the scale $\mu = m_{\tilde t}$, improved by the 
NLO K-factor~\footnote{Although gluon radiation at NLO leads to a 
small shift in the stop $p_T$ distribution
to lower $p_T$ values~\cite{PROSPINO,PROSPINOstops}, 
we do not expect this shift to affect our analysis
in any significant way.} 
obtained from PROSPINO \cite{PROSPINO,PROSPINOstops} 
(see Fig.~\ref{fig:xsecmass}).
The K-factor varies between 1 and 1.5 for $m_{\tilde t}$ 
decreasing from 450 to 100 GeV.
We use the CTEQ5 parton distribution functions \cite{CTEQ5}.
We assume that the gluino and the other squarks are heavy enough that
they do not affect the NLO cross section.  This is already the case
for gluino and squark masses above about 200 GeV \cite{PROSPINOstops}.

Top squarks can also be produced via cascade decays of heavier
supersymmetric particles, with a highly model dependent rate.
To be conservative, we assume that the masses of the heavier 
supersymmetric particles are large enough that their production
rate at Tevatron energies can be neglected.

%------------------------------------------------------------------
\section{Previous studies of stops at Run II}
\label{sec:literature}

A number of previous studies have considered the prospects for stop
discovery at Run~II of the Tevatron, which we summarize here.
In general, the most detailed SUSY studies have been done in the context of 
supergravity; in this case SUSY is broken at the Planck 
scale so that the gravitino
plays no role in the collider phenomenology.  Then the lightest 
neutralino is the LSP and ends all superparticle decay chains.  
The reach of the Tevatron for a number of stop decay modes 
has been analyzed in Refs.~\cite{Baer,Demina}.  
The signal depends on the 
decay chain, which in turn depends on the relative masses of 
various SUSY particles.
For sufficiently heavy stops, the decay $\tilde t \to t \tilde \chi_1^0$
will dominate.  This channel is of limited use at Run II because 
the stop pair production cross section falls rapidly with increasing stop mass,
and this channel requires $m_{\tilde t} > m_t + m_{\tilde \chi_1^0}$,
which is quite heavy for the Tevatron in the case of minimal 
supergravity.\footnote{In low-energy SUSY breaking scenarios, however, 
the mass range $m_{\tilde t} > m_t + m_{\tilde \chi_1^0}$ is interesting
at Tevatron energies because the distinctive signal allows backgrounds
to be reduced to a very low level, as we will show.}
For lighter stops, if a chargino is lighter than the stop then 
$\tilde t \to b \tilde \chi_1^+$ tends to dominate (followed by the
decay of the chargino).
The details of the signal depend on the Higgsino content of $\tilde \chi_1^+$.
If $\tilde \chi_1^+$ has a mass larger than $m_{\tilde{t}} - m_b$,
the previous decay does not occur and the three-body decay
$\tilde t \to b W^+ \tilde \chi_1^0$ dominates; this decay proceeds through
the exchange of a virtual top quark, chargino, or bottom squark.
If the stop is too light to decay into
an on-shell $W$ boson and $\tilde \chi_1^0$, then the flavor-changing decay
$\tilde t \to c \tilde \chi_1^0$ tends to dominate.
Finally, if a sneutrino or slepton is light, then 
$\tilde t \to b \ell^+ \tilde \nu_{\ell}$ or 
$\tilde t \to b \tilde \ell^+ \nu_{\ell}$, respectively, will occur
(followed by the decays of the slepton or the sneutrino, if it is not the LSP).
At Run~II with 2 (20) fb$^{-1}$ of total integrated luminosity,
in the context of minimal supergravity
one can probe stop masses up to 160 (200) GeV 
in the case of the flavor changing decay, while 
stop masses as high as 185 (260) GeV can be 
probed if the stop decays into a bottom quark and 
a chargino~\cite{Demina}.
A similar reach holds in the case of a light sneutrino~\cite{Demina}
and in the case of large $\tan\beta$ when the stop can decay into
$b \tau \nu \tilde \chi_1^0$ \cite{Djouadi}.

One can also search for stops in the decay products of other SUSY 
particles \cite{Demina}.
In top decays, the process $t \to \tilde t \tilde g$ is already 
excluded because of the existing lower bound on the gluino 
mass~\footnote{The bound on the gluino mass is, however, model
dependent. Under certain conditions, an allowed window exists
for gluino masses below the gauge boson masses~\cite{stuart,lightsb}.}.
Other possibilities are $\tilde \chi^- \to b \tilde t^*$ and
$\tilde g \to t \tilde t^*$.
Finally, the decays $\tilde b \to \tilde t W^-$ and $\tilde t H^-$ have 
to compete with the preferred decay, $\tilde b \to b \tilde \chi_1^0$, 
and so may have small branching ratios 
(depending on the masses of $\tilde t$, $\tilde b$, 
$\tilde \chi_1^0$ and $H^-$).  The signals for these processes at the 
Tevatron Run~II have been considered in minimal supergravity in 
Ref.~\cite{Demina}.

Relatively few studies have been done in the context of low-energy 
SUSY breaking with a gravitino LSP.
A study of GMSB signals performed as part of the Tevatron Run~II 
workshop \cite{SUSYRun2} considered the decays of various SUSY
particles as the NLSP.  As discussed before, the NLSP in such models
will decay directly to the gravitino and Standard Model particles.
If the stop is the NLSP in such a model, then
it will decay via $\tilde t \to t^{(*)} \tilde G \to b W^+ \tilde G$
(for $m_{\tilde t} > m_b + m_W$).
Note that because $\tilde G$ is typically very light in such models, 
$m_{\tilde t} > m_W + m_b$ is sufficient for this decay to proceed
with an on-shell $W$ boson.
Ref.~\cite{Demina} found sensitivity at Run~II to this decay mode for
stop masses up to 180 GeV with 4 fb$^{-1}$.
This stop decay looks very much like
a top quark decay; nevertheless, even for stop masses near $m_t$, 
such stop decays can be separated from top quark decays
at the Tevatron using kinematic correlations among the decay 
products \cite{Peskin}.

Finally, Ref.~\cite{KaneTev} considered general GMSB signals at the Tevatron
of the form $\gamma \gamma \missET + X$. The authors of
Ref.~\cite{KaneTev} provide an analysis of the possible bounds on 
the stop mass coming from the Run I Tevatron data. They 
analyze the stop decay mode into a charm quark and a neutralino, and also
possible three body decays, by scanning over a sample of models. 
They conclude that stop masses smaller than 140 GeV can be
excluded already by the Run I Tevatron data within low energy supersymmetry
breaking models independent of the stop decay mode, 
assuming that $m_{\tilde \chi_1^0} > 70$ GeV.

%------------------------------------------------------------------
\section{Top squark decay branching ratios}
\label{sec:stopBRs}

The decay properties of the lighter stop depend on the supersymmetric
particle spectrum.  Of particular relevance are the mass splittings
between the stop and the lightest chargino, neutralino and bottom squark.
In our analysis, we assume that the charginos and bottom squarks are
heavier than the lighter stop, so that the on-shell decays 
$\tilde t \to \tilde b W$
and $\tilde t \to \tilde \chi^+ b$ are kinematically forbidden.
Then the details of the stop decay depend on the mass splitting between
the stop and the lightest neutralino.
If $m_{\tilde t} < m_W + m_b + m_{\tilde \chi_1^0}$, two decay modes
are kinematically accessible: 
\begin{enumerate}
\item the flavor-changing (FC) two-body decay 
$\tilde t \to c \tilde \chi^0_1$. 
This two-body decay  proceeds through a flavor-changing loop
involving $W^+$, $H^+$ or $\tilde \chi^+$ exchange or through a tree-level
diagram with a $\tilde t - \tilde c$ mixing mass insertion;
\item the four-body decay via a virtual $W$ boson,
$\tilde t \to W^{+*} b \tilde \chi_1^0 \to j j b \tilde \chi_1^0$ or 
$\ell \nu b \tilde \chi_1^0$~\cite{4body}. 
\end{enumerate}
The branching ratio of the stop decay into charm and neutralino 
strongly depends on the details of the supersymmetry breaking mechanism.
In models with no flavor violation at the messenger scale, the whole
effect is induced by loop effects and receives a logarithmic enhancement
which becomes more relevant for larger values of the messenger mass.
In the minimal supergravity case, the two-body FC decay branching ratio tends
to be the dominant one. In the case of low energy supersymmetry
breaking, this is not necessarily the case.  Since the analysis
of the four-body decay process is very similar to the three-body decay
described below for larger mass splittings between the stop and
the lighter neutralino, here we shall
analyze only the case in which the two-body FC decay 
$\tilde t \to c \tilde \chi^0_1$ is the dominant one
whenever $m_{\tilde t} < m_W + m_b + m_{\tilde \chi_1^0}$.

For larger mass splittings, so that 
$m_W + m_b + m_{\tilde \chi^0_1} < m_{\tilde t} < 
m_t + m_{\tilde \chi^0_1}$,
the three-body decay
$\tilde t \to W^+ b \tilde \chi_1^0$ becomes accessible, with
$\tilde \chi^0_1 \to \gamma \tilde G$.  
This stop decay
proceeds through a virtual top quark, virtual charginos, or virtual 
sbottoms.  Quite generally, this 3-body decay will dominate over
the 2-body FC decay in this region of phase space.

For still heavier stops,
$m_{\tilde t} > m_t + m_{\tilde \chi_1^0}$, 
the two-body tree-level decay mode 
$\tilde t \to t \tilde \chi_1^0$ becomes kinematically accessible
and will dominate. Although
the 3-body and 2-body FC decays are still present, their
branching ratios are strongly suppressed.

\begin{figure}
\resizebox{\textwidth}{!}{
\rotatebox{270}{\includegraphics{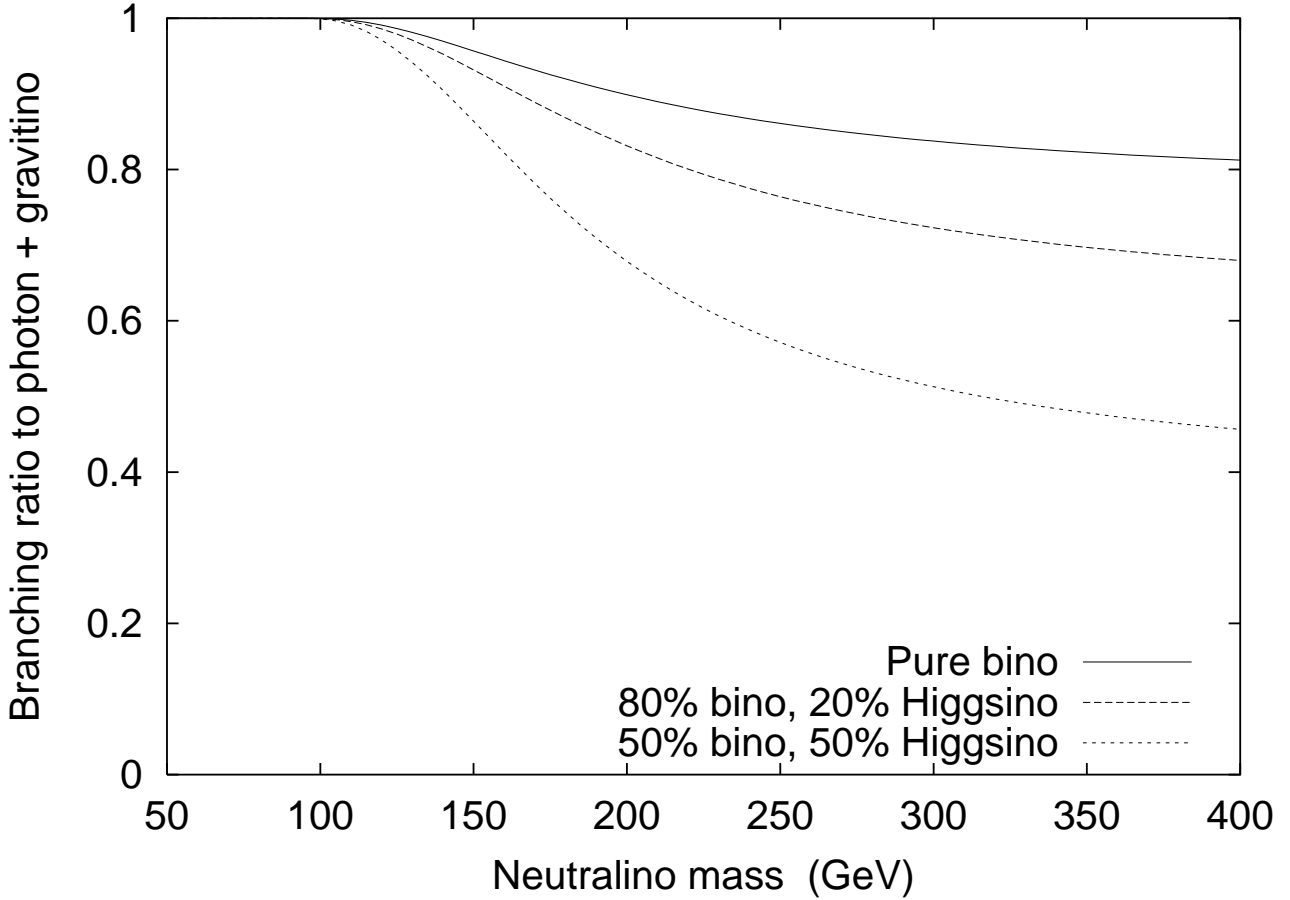}}
}
\caption{Branching ratio for the decay of the lightest neutralino into
a photon and a gravitino as a function of the neutralino mass.
Shown are the branching ratio if the neutralino is a pure bino
(solid line) and for 20\% and 50\% Higgsino admixtures (long and 
short dashed lines, respectively) assuming that $m_{h^0} = 120$ GeV
and that the other MSSM Higgs bosons are very heavy.  
For the Higgsino admixture in the lightest neutralino,
we choose the $\tilde H_1$--$\tilde H_2$ mixing so that the field 
content is aligned with that of $h^0$ and the longitudinal component
of the $Z$ boson in order to minimize the branching ratio to photons.
}
\label{fig:binoBR}
\end{figure}

Let us emphasize that, since the bino is an admixture of the zino
and the photino, a pure bino neutralino can decay into either
$\gamma \tilde G$ or $Z \tilde G$ (see Eq.~\ref{rategm}).
If the lightest neutralino is a mixture of bino and wino components,
then the relative zino and photino components can be varied arbitrarily, 
leading to a change in the relative branching ratios to $\gamma \tilde G$ and 
$Z \tilde G$.
If the lightest neutralino contains a Higgsino component, then the
decay to $h^0 \tilde G$ is also allowed.
We show in Fig.~\ref{fig:binoBR} the branching ratio of the lightest
neutralino into $\gamma \tilde G$ as a function of its mass and
Higgsino content.

%------------------------------------------------------------------
\section{Top squark signals in low-energy SUSY breaking}
\label{sec:signals}

As discussed in the previous section,
the decay properties of the lighter stop depend primarily on 
the mass splitting between the stop and the lightest neutralino.
In this section, we proceed with the phenomenological 
analysis of the signatures of top squark production associated 
with the different decay modes. In Sec.~\ref{sec:2body} we
shall analyze the signatures associated with the two-body FC 
decay, which after the neutralino decay leads
to  $\tilde t \to c \gamma \tilde G$. In Sec.~\ref{sec:3body}
we shall analyze the signatures associated with the three-body
decay, which after neutralino decay leads to 
$\tilde t \to b W^+ \gamma \tilde G$. The
two-body decay $\tilde t \to t \tilde \chi_1^0$, which typically dominates
for $m_{\tilde t} > m_t + m_{\tilde \chi_1^0}$, leads to the same
final state as the three-body
decay and therefore the discussion of this case will be
included in Sec.~\ref{sec:3body}.
Finally, in Sec.~\ref{sec:topdecay} we consider the possibility that stops
are produced in the decays of top quarks.

%----------------------------------------------------------------------
\subsection{Two-body FC decay: $\tilde t \to c \gamma \tilde G$}
\label{sec:2body}

With the stop undergoing the aforementioned decay, the final state consists
of a pair each of charm jets, photons and (invisible) gravitinos. 
As we will show below, the backgrounds to this process are small enough 
that we need not require charm tagging.
Thus the signal consists of
\[
         2 ({\rm jets}) + 2 \gamma + \missET.
\]
The selection criteria we adopt are:
\begin{enumerate}
        \item each event must contain two jets and two photons, each of which
              should 
              have a minimum transverse momentum ($p_T > 20$ GeV) and 
              be contained in the pseudorapidity range $-2.5 < \eta < 2.5$;
        \item the jets and the photons should be well separated from 
              each other; namely,
              \[ 
                \delta R_{j j } > 0.7, \quad
                \delta R_{\gamma \gamma} > 0.3, \quad
                \delta R_{j \gamma } > 0.5
              \]
        where $\delta R^2 = \delta \eta^2 + \delta \phi^2$, with 
        $\delta \eta$ ($\delta\phi$) denoting the difference in 
        pseudorapidity (azimuthal angle) of the two 
        entities under consideration;
        \item the invariant mass of the two jets should be sufficiently 
                far away from the $W$- and $Z$-masses:\\ 
                $m_{j j} \notin$ (75 GeV, 95 GeV);
        \item each event should 
              be associated with a minimum missing transverse 
              momentum ($\misspT > 30$ GeV).
\end{enumerate}
The photons and jets in signal events tend to be very central; in particular,
reducing the pseudorapidity cut for the two photons to
$-2.0 < \eta < 2.0$ would reduce the signal by less than about 3\%.
(This change would reduce the background by a somewhat larger fraction.)
Apart from ensuring observability, these selection criteria also serve to 
eliminate most of the backgrounds, which are listed in Table \ref{tab:2bodybg}.

\begin{table}
\begin{center}
\begin{tabular}{|l|r|r|}
\hline
Background & Cross section after cuts & after $\gamma$ ID \\
\hline \hline
$jj\gamma\gamma Z$, $Z \to \nu \bar \nu$ & $\sim 0.2$ fb & $\sim 0.13$ fb \\
$j j \gamma \nu \bar \nu$ + $\gamma$ radiation& $\sim 0.002$ fb & $\sim 0.001$ fb\\
\hline
$b \bar b \gamma\gamma$, $c \bar c \gamma\gamma$ & $\lsim 0.1$ fb 
        & $\sim 0.06$ fb\\
\hline
$jj \gamma \gamma$ & $\sim 0.2$ fb & $\sim 0.13$ fb\\
\hline \hline
Backgrounds with fake photons: & & \\
\hline
$jj(ee \to \gamma\gamma)$ & $\sim 5 \times 10^{-4}$ fb 
        & $\sim 5 \times 10^{-4}$ fb \\
\hline
$jj\gamma(j \to \gamma)$ & $\sim 0.8$ fb & $\sim 0.8$ fb\\
\hline
$jj(jj \to \gamma\gamma)$ & $\sim 0.8$ fb & $\sim 0.8$ fb\\
\hline \hline
Total  & $\sim 2$ fb & $\sim 2$ fb \\
\hline
\end{tabular}
\end{center}
\caption{Backgrounds to $\tilde t \tilde t^* \to jj\gamma\gamma \missET$.
The photon identification efficiency is taken to be $\epsilon_{\gamma} = 0.8$
for each real photon.  See text for details.}
\label{tab:2bodybg}
\end{table}
 
A primary source of background is the SM production of $j j \gamma \gamma
\nu_i \bar \nu_i$ where the jets could have arisen from either quarks or 
gluons in the final state of partonic subprocesses. A full diagrammatic 
calculation would be very computer-intensive and is beyond the scope of 
this work. Instead, we consider the subprocesses that are expected 
to contribute the bulk of this particular background, namely 
$p \bar p \rightarrow 2 j + 2 \gamma + Z + X$ with the $Z$ subsequently 
decaying into neutrinos. These processes are quite tractable and 
were calculated with the aid of the helicity amplitude program 
{\sc Madgraph}~\cite{Madgraph}. On imposition of the abovementioned set of 
cuts, this background at the Run II falls to below $\sim 0.2$ fb.

An independent estimate of the $j j \gamma \gamma \nu_i \bar \nu_i$
background may be obtained through the consideration of 
the single-photon variant, namely $j j \gamma \nu_i \bar \nu_i$ production, 
a process that {\sc Madgraph} can handle.  After imposing the same 
kinematic cuts (other than requiring only one photon) as above, this process 
leads to a cross section of roughly 0.2 fb.  Since the emission of a 
second hard photon should cost us a further power of $\alpha_{\rm em}$, the 
electromagnetic coupling constant, this background falls to innocuous levels.
We include both this estimate and the one based on $Z$ production from
the previous paragraph in Table~\ref{tab:2bodybg}. While a naive addition
of both runs the danger of overcounting, this is hardly of any importance 
given the overwhelming dominance of one.

A second source of background is $b\bar b\gamma\gamma$ 
($c \bar c\gamma\gamma$),
with missing transverse energy coming from the semileptonic 
decay of one or both of the $b$ ($c$) mesons.
In this case, though, the neutrinos tend to be soft
due to the smallness of the $b$ and $c$ masses. 
Consequently, the  cut on $\misspT$ serves to eliminate most of 
this background, leaving behind less than 0.1 fb. 
This could be further reduced (to $\lsim 0.001$ fb) 
by vetoing events with leptons in association with jets. 
However, such a lepton veto would significantly impact the selection
efficiency of our signal, which contains $c \bar c$, thereby reducing
our overall sensitivity in this channel.

A third source of background is
$j j \gamma \gamma$ production, 
in which the jet (light quark or gluon) and/or photon energies 
are mismeasured leading to a fake $\misspT$.
To simulate the effect of experimental resolution, we use a
(very pessimistic) Gaussian 
smearing: $\delta E_j / E_j = 0.1 + 0.6 / \sqrt{E_j ({\rm GeV})}$
for the jets and  
$\delta E_\gamma / E_\gamma = 0.05 + 0.3 / \sqrt{E_\gamma ({\rm GeV})}$
for the 
photons.\footnote{We have also performed similar smearing for the other 
        background channels as well as for the signal. However, there 
        it hardly is of any importance as far as the estimation 
        of the total cross section is concerned.} 
While the production cross section is much higher than that of any
of the other backgrounds considered, the ensuing 
missing momentum tends to be small; in particular, our cut on
$\missET$ reduces this background by almost 99\%.\footnote{This
reduction factor is in rough agreement with that found for
$\gamma + j$ events in CDF Run I data in Ref.~\cite{gammabCDF}.}
On imposition of our cuts, this fake background is reduced to 
$\sim 0.2$ fb.\footnote{We note that in the squark searches in supergravity 
scenarios the signal is $j j \missET$; in this case a
similar background due to dijet production with fake $\misspT$ is
present.  This background is larger by two powers of $\alpha$ than
the $jj\gamma\gamma$ background, yet it can still be reduced to an
acceptable level by a relatively hard cut on $\misspT$ 
(see, {\it e.g.}, Ref.~\cite{Demina}).}  

Finally, we consider the instrumental backgrounds from electrons or
jets misidentified as photons.  Based on Run I analyses \cite{RunIee}
of electron pair production and taking the probability for an 
electron to fake a photon to be about 0.4\%, we estimate the
background due to electrons faking photons to be of order 
$5 \times 10^{-4}$ fb.
More important is the background in which a jet fakes a photon.
Based on a Run I analysis \cite{gammabCDF} and taking the probability
for a jet to fake a photon to be about 0.1\%, we estimate the background
due to $jjj\gamma$ in which one of the jets fakes a second photon to
be about 0.8 fb.  
Similarly, we estimate that the background due to $jjjj$ in which two of the 
jets fake photons is somewhat smaller; to be conservative we take it 
to be of the same order, {\it i.e.}, 0.8 fb.

\begin{figure}
\resizebox{\textwidth}{!}{
\includegraphics*[0,0][550,550]{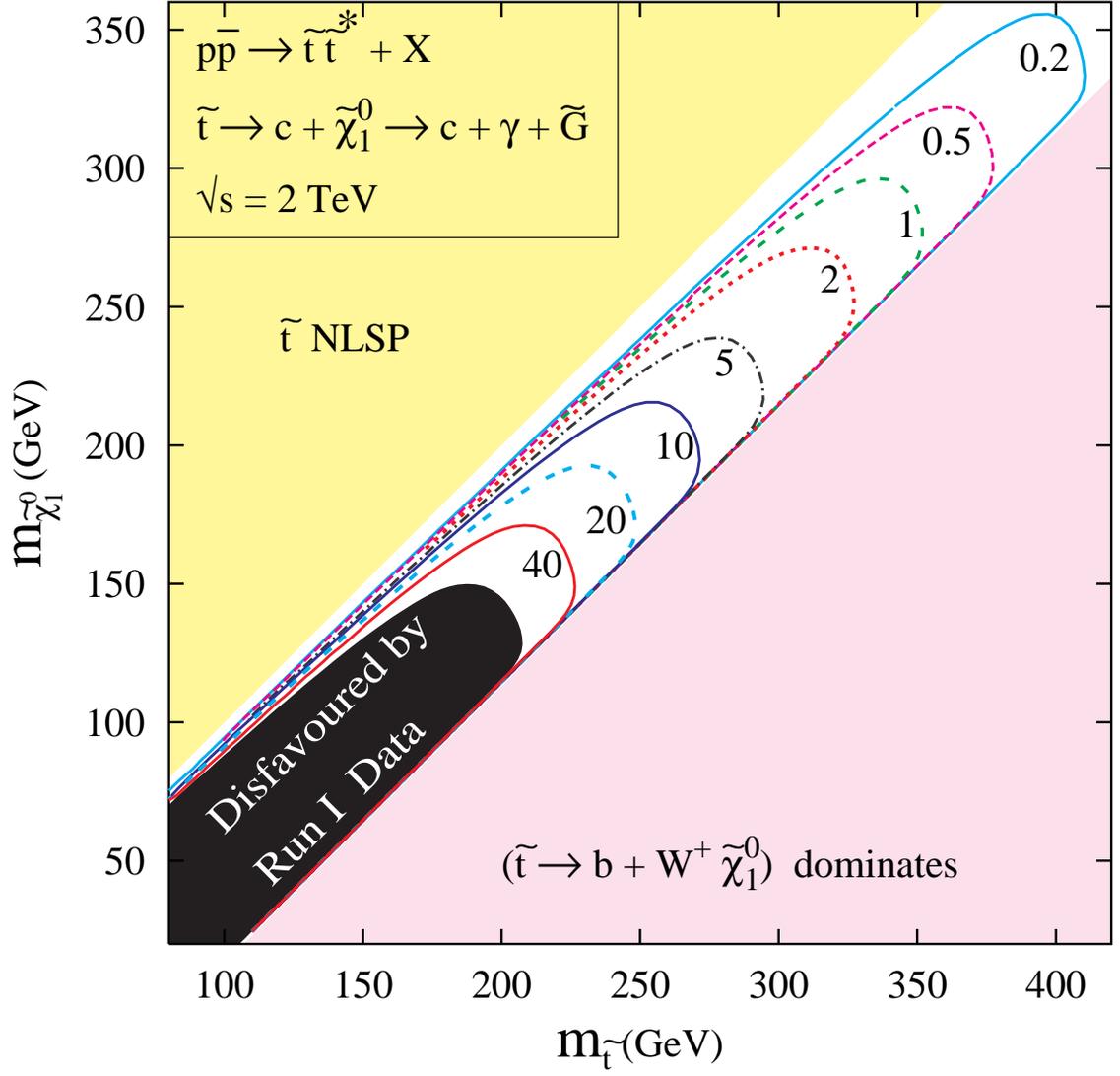}
}
\caption{Cross sections in fb for stop pair production in Run II with 
$\tilde t \to c \gamma \tilde G$, after cuts.  No efficiencies have
yet been applied.  The black area is 
excluded by non-observation of $jj\gamma\gamma\missET$ events in Run I.
}
\label{fig:2bodyxsec}
\end{figure}

Having established that the backgrounds are small, let us now turn to 
the signal cross section that survives the cuts. 
In Fig.~\ref{fig:2bodyxsec}, we present these as contours 
in the $m_{\tilde t}$--$m_{\tilde \chi_1^0}$ plane.
We assume that the branching ratio of $\tilde t \to c \tilde \chi_1^0$ 
dominates in the region of parameter space that we consider here.  
The branching ratio of $\tilde \chi_1^0 \to \gamma \tilde G$ is
taken from Fig.~\ref{fig:binoBR} assuming that $\tilde \chi_1^0$ is 
a pure bino.\footnote{If $\tilde \chi_1^0$ is not a pure bino,
its branching ratio to $\gamma\tilde G$ will typically be reduced
somewhat when its mass is large (see Fig.~\ref{fig:binoBR}).
This will lead to a reduction of the signal cross section at 
large $m_{\tilde \chi_1^0}$ by typically a few tens of percent.}
All our plots are made before detector efficiencies are applied.
With a real detector, each photon is identified with about 
$\epsilon_{\gamma} = 80$\% efficiency;
thus the cross sections shown in Fig.~\ref{fig:2bodyxsec} must be 
multiplied by $\epsilon_{\gamma}^2 = 0.64$ in order to obtain numbers of
events.\footnote{The diphoton trigger efficiency is close to 100\%,
so we neglect it here.}
While the production cross sections 
are independent of the neutralino mass, the decay kinematics 
have a strong dependence on $m_{\tilde \chi_1^0}$. 
For a given $m_{\tilde t}$, a small mass splitting between $m_{\tilde t}$
and $m_{\tilde \chi_1^0}$ would imply a soft charm jet, which would 
often fail to satisfy our selection criteria.  This causes the gap between
the cross section contours and the 
upper edge of the parameter space band that we are 
exploring in Fig.~\ref{fig:2bodyxsec}.  This is 
further compounded at large $m_{\tilde \chi_1^0}$ 
by the fact that a large neutralino mass typically implies 
a smaller $\tilde \chi_1^0 \to \gamma \tilde G$ branching fraction 
(see Fig.\ref{fig:binoBR}).  On the other hand, a 
small $m_{\tilde \chi_1^0}$ results in reduced momenta for the 
gravitino and the photon, once again resulting in a loss of signal;
however, this is important only for $m_{\tilde \chi_1^0} \lsim 70$ GeV, 
which is not relevant in this search channel.

The signal cross sections are fairly substantial.  In particular,
the dark area in Fig.~\ref{fig:2bodyxsec}
corresponds to a signal cross section of 50 fb or larger
at Run I of the Tevatron.
Taking into account the identification efficiency of 64\% for the 
two photons, such a cross section would have yielded 3 signal events in the 
100 pb$^{-1}$ collected in Run I over a background of much less 
than 1 event.  Run I can thus exclude this region at 95\% confidence level.
In particular, we conclude that 
Run I data excludes stop masses up to about 200 GeV,
for large enough mass splitting between the stop and the neutralino.
When the stop-neutralino mass splitting is small ({\it i.e.}, less than
about $10-20$ GeV), the charm quark jets 
become too soft and the signal efficiency decreases dramatically.
For comparison, a D{\O} search \cite{DZerosquark} for inclusive
$p \bar p \to \tilde \chi^0_2 + X$ with 
$\tilde \chi_2^0 \to \gamma \tilde \chi_1^0$
in the context of minimal supergravity yields a limit on the 
production cross section of about 1 pb for parent squark masses of
order 150-200 GeV.  Interpreting this in terms of stop pair production 
with $\tilde \chi_2^0 \to \gamma \tilde \chi_1^0$ reidentified as 
$\tilde \chi_1^0 \to \gamma \tilde G$ increases the signal efficiency
by a factor of $\sim 2.7$ because every SUSY event now contains two 
photons \cite{DZerosquark}; the D{\O} analysis then
yields a stop mass bound of about 180 GeV, in rough agreement with
our result.\footnote{Note, however, that the non-negligible 
mass of the LSP of about 35 GeV assumed in 
Ref.~\cite{DZerosquark} leads to kinematics that differ significantly 
from those in our analysis, in which the gravitino is essentially massless.}
In addition, Ref.~\cite{KaneTev} projected a Run I exclusion of stops
in this decay channel for masses below about 160-170 GeV, again in rough
agreement with our result.

To claim a discovery at the $5\sigma$ level, 
one must observe a large enough number of events that the 
probability for the background to fluctuate up to that level
is less than $5.7 \times 10^{-7}$.  Because the number of expected
background events in this analysis is small, we use Poisson statistics
to find the number of signal events required for a $5\sigma$ discovery.
Taking the total background cross section to be 2 fb from Table 
\ref{tab:2bodybg},
we show in Table~\ref{tab:2bodyevents} the expected maximum stop 
discovery mass reach at Tevatron Run~II for various amounts of integrated
luminosity.\footnote{The cross sections and numbers of events required
for discovery are quoted in terms of integrated luminosities at a 
single detector.  
If data from the CDF and D{\O} detectors are combined,
the integrated luminosity of the machine is effectively doubled.}
In particular, with 4 fb$^{-1}$ a stop discovery can
be expected in this channel if $m_{\tilde t} < 285$ GeV, 
with $S/B$ of more than 2/1.\footnote{For comparison, in the case of minimal
supergravity a reach of $m_{\tilde t} < 180$ GeV can be expected
in the $\tilde t \to c \tilde \chi_1^0$ channel
with 4 fb$^{-1}$ at Tevatron Run II; the same reach is 
obtained in low-energy SUSY breaking scenarios in which the stop is the NLSP
rather than the neutralino~\cite{Demina}.  
In both of these cases, the signal consists
of $jj\missET$, with no photons in the final state.}
Including the effects of mixing in the composition of the lightest neutralino,
a 50\% reduction in the $\tilde \chi_1^0 \to \gamma \tilde G$ branching
ratio compared to the pure bino case would reduce the 
stop mass reach by only about 20 GeV.  
For such a reduction to occur in the relevant neutralino mass range
of about $200 - 250$ GeV, the lightest neutralino would have to be 
less than half bino.

\begin{table}
\begin{center}
\begin{tabular}{|c|c|c|c|c|}
\hline
$\int\mathcal{L}$ & B & S for a 5$\sigma$ discovery 
& $\sigma_S \times \epsilon_{\gamma}^2$
   & Maximum stop mass reach \\
\hline
2 fb$^{-1}$  &  4 & 14 &  7.0  fb & 265 GeV\\
4 fb$^{-1}$  &  8 & 18 &  4.5 fb & 285 GeV\\
15 fb$^{-1}$ & 30 & 31 &  2.1 fb & 310 GeV\\
30 fb$^{-1}$ & 60 & 42 &  1.4 fb & 325 GeV\\
\hline
\end{tabular}
\end{center}
\caption{
Number of signal events (S) required for a 5$\sigma$ stop discovery at 
Tevatron Run~II in the $jj\gamma\gamma \missET$ channel
and the corresponding signal cross section after cuts and efficiencies
and maximum stop mass reach.  We assume a photon identification efficiency
of $\epsilon_{\gamma} = 0.80$.
The number of background events (B) is based on a background cross section 
of 2 fb from Table~\ref{tab:2bodybg}.
}
\label{tab:2bodyevents}
\end{table}

%-----------------------------------------------------------------------
\subsection{Three-body decay: $\tilde t \to b W^+ \gamma \tilde G$}
\label{sec:3body}

For a large enough splitting between the stop and neutralino 
masses, the signature of stop pair production would 
be\footnote{The backgrounds are again small enough
after cuts in this channel that we do not need to tag the $b$ quarks.
In the case of a discovery, one could imagine tagging the $b$ quarks and 
reconstructing the $W$ bosons in order to help identify the discovered
particle.} $jjWW\gamma\gamma\missET$.  In this analysis, we consider 
only the dominant hadronic decay mode of both $W$ bosons.

This decay mode of the stop proceeds via three Feynman diagrams,
involving an intermediate off-shell top quark, chargino or sbottom.  
We use the full decay matrix elements as given in Ref.~\cite{Porod}.
Although this introduces several additional parameters into the analysis,
a few simplifying assumptions may be made without becoming too model 
dependent. For example, assuming that the lightest stop is predominantly 
the superpartner of the right-handed top-quark ($\tilde t_R$), 
eliminates the sbottom exchange diagram altogether. 
Even if the stop contains a mixture of $\tilde t_R$ and $\tilde t_L$,
under our assumption that the lighter stop is the 
next-to-lightest Standard Model superpartner, the sbottom exchange
diagram will be suppressed by the necessarily larger sbottom mass.
As for the chargino exchange, the wino component does not contribute 
for a $\tilde t_R$ decay.  Thus the chargino contribution is dominated
by its Higgsino component.  Furthermore, if we  
concentrate on scenarios with large 
values of the supersymmetric mass parameter $\mu$ and of the wino mass
parameter (in which case the charginos are heavy and
the neutralino is almost a pure bino),   the chargino exchange
contribution is also suppressed and the dominant decay mode is
via the diagram involving an off-shell top quark.  
To simplify our numerical calculations, we have then assumed
that only this diagram contributes to the stop decay matrix element.
We have checked for a few representative points though, 
that the inclusion of the sbottom and chargino diagrams 
does not significantly change the signal efficiency after cuts as long as 
we require that $m_{\tilde b}$, $m_{\tilde \chi^+} > m_{\tilde t}$.

As the $W$ bosons themselves decay, it might be argued that their 
polarization information needs to be retained. 
However, since we do not consider angular correlations between the decay 
products, this is not a crucial issue; the loss 
of such information at intermediate steps in the decay does not lead to a 
significant change in the signal efficiency after cuts.  
This is particularly true for the hadronic 
decay modes of the $W$, for which the profusion of jets frequently leads to
jet overlap, thereby obscuring detailed angular correlations.  
It is thus safe to make the approximation of neglecting the polarization
of the $W$ bosons in their decay distributions, and we do so in our analysis.

Before we discuss the signal profile and the backgrounds, let us elaborate 
on the aforementioned jet overlapping. With six quarks in the final state, 
some of the resultant jets will very often be too close to each 
other to be recognizable as coming from different partons.
We simulate this as follows.  We count a final-state parton (quark or gluon) 
as a jet
only if it has a minimum energy of 5 GeV and lies within the pseudorapidity
range $-3 < \eta < 3$.  We then
merge any two jets that fall within a $\delta R$ separation of 0.5;
the momentum of the resultant jet is the sum of the two momenta. 
We repeat this process iteratively, starting with the hardest jet.
Our selection cuts are then applied to the (merged) jets that survive 
this algorithm. 

The signal thus consists of:
\[
 n \; {\rm jets} +  2 \gamma + \missET  \qquad (n \leq 6)
\]
Hence, all of the SM processes discussed in the previous section 
yield backgrounds to this signal when up to four additional jets are radiated. 
Now, the radiation of each hard and well separated jet 
suppresses the cross section by a factor of order $\alpha_s \simeq 0.118$.  
Then, since the $jj\gamma\gamma \missET$ backgrounds are already quite 
small after the cuts applied in the previous section 
(see Table~\ref{tab:2bodybg}), the backgrounds with additional jets 
are expected to be still smaller.~\footnote{For the backgrounds in 
which one or both of the photons are faked by misidentified jets,
we have taken into account the larger combinatoric factor that
arises when more jets are present to be misidentified.}

There exists a potential exception to the last assertion, 
namely the background due to $t \bar t \gamma \gamma$ production. 
To get an order of magnitude estimate,  the 
cross section for $t \bar t$
production at Tevatron Run~II is 8 pb \cite{topxsec}. If both of
the photons are required to be energetic and isolated, we would expect a 
suppression by a factor of order $\alpha^2_{\rm em}$, 
leading to a cross section 
of the order of 0.5 fb.  This cross section is large enough that we 
need to consider it carefully. 

If both $W$ bosons were to decay hadronically, 
then a sufficiently large missing transverse energy 
can only come from a mismeasurement of the jet or
photon energies. As we have seen in the previous section, this 
missing energy is normally too small to pass our cuts, 
thereby suppressing the background. 
If one of the $W$ bosons decays leptonically, however, it will yield 
a sizable amount of $\missET$.  This background can be largely eliminated
by requiring that no lepton ($e$ or $\mu$) is seen in the detector.
This effectively eliminates the $W$ decays into $e$ or $\mu$ or decays
to $\tau$ followed by leptonic $\tau$ decays; the remaining background
with hadronic $\tau$ decays is naturally quite small without requiring
additional cuts. Considerations such as these lead us to an 
appropriate choice of criteria  for an event to be selected:
\begin{enumerate}
\item At least four jets, each with a minimum transverse momentum 
      $p_{Tj} > 20$ GeV and
      contained in the pseudorapidity interval of $-3 < \eta_j < 3$.
      Any two jets must be separated by $\delta R_{jj} > 0.7$. 
      As most of the signal events do end up with 4 or more energetic 
      jets (the hardest jets coming typically from the $W$ boson decays), 
      this does not cost us in terms of the signal, while reducing the 
      QCD background significantly. In addition, the $t \bar t \gamma \gamma$ 
      events with both $W$'s decaying leptonically are reduced to a 
      level of order $10^{-4}$ fb by this requirement alone.
        
\item Two photons, each with $p_{T\gamma} > 20$ GeV 
      and pseudorapidity $-2.5 < \eta_\gamma < 2.5$.  
      The two photons must be separated by at least 
        $\delta R_{\gamma \gamma} > 0.3$.

\item Any photon-jet pair must have a minimum separation of 
        $\delta R_{j \gamma} > 0.5$.

\item A minimum missing transverse energy $\missET > 30$ GeV.

\item The event should not contain any isolated 
      lepton with $p_T > 10$ GeV
      and lying within the pseudorapidity range $-3 < \eta < 3$.
\end{enumerate}

As in the previous section, the cut on $\missET$ serves to eliminate 
most of the background events with only 
a fake missing transverse momentum (arising out of 
mismeasurement of jet energies).  In association with the lepton 
veto, it also eliminates the bulk of events in which one of the $W$
bosons decays leptonically (including the $\tau$ channel). 
A perusal of Table~\ref{tab:3bodybg}, which summarizes the
major backgrounds after cuts, convinces us that the backgrounds to
this channel are very small, in fact much smaller than those for the 
previous channel. 

\begin{table}
\begin{center}
\begin{tabular}{|l|r|r|}
\hline
Background & Cross section after cuts & after $\gamma$ ID \\
\hline \hline
$(jj\gamma\gamma Z$, $Z \to \nu \bar \nu) + 2j$ & $\sim 0.003$ fb
        & $\sim 0.002$ fb \\
$j j \nu \bar \nu \gamma \gamma +2j$ & $\sim 3 \times 10^{-5}$ fb
        & $\sim 2 \times 10^{-5}$ fb\\
\hline
$(b \bar b \gamma\gamma$, $c \bar c \gamma\gamma) + 2j$ & $\sim 0.001$ fb
        & $\sim 0.0006$ fb\\
\hline
$jj \gamma \gamma + 2j$ & $\lsim 0.003$ fb & $\lsim 0.002$ fb\\
\hline
$t \bar t \gamma\gamma$, $WW \to jjjj$ & $\lsim 10^{-4}$ fb 
        & $\lsim 10^{-4}$ fb \\
$t \bar t \gamma\gamma$, $WW \to jj \ell \nu$, $\ell = e$, $\mu$, or 
$\tau \to \ell$
        & $\sim 0.001$ fb & $\sim 0.0006$ fb\\
$t \bar t \gamma\gamma$, $WW \to jj \tau \nu$, $\tau \to j$ 
        & $\lsim 0.01$ fb & $\lsim 0.006$ fb\\
\hline \hline
Backgrounds with fake photons: & & \\
\hline
$jj(ee \to \gamma\gamma) + 2j$ & $\sim 7\times 10^{-6}$ fb 
        & $\sim 7\times 10^{-6}$ fb\\
\hline
$jj\gamma(j \to \gamma) + 2j$ & $\sim 0.02$ fb & $\sim 0.02$ fb\\
\hline
$jj(jj \to \gamma\gamma) + 2j$ & $\sim 0.03$ fb & $\sim 0.03$ fb\\
\hline \hline
Total  &  $\lsim 0.07$ fb & $\sim 0.06$ fb\\
\hline
\end{tabular}
\end{center}
\caption{
Backgrounds to $\tilde t \tilde t^* \to jjWW\gamma\gamma\missET$ with
both $W$ bosons decaying hadronically.  The photon identification
efficiency is taken to be $\epsilon_{\gamma} = 0.8$ for each
real photon.  See text for details.}
\label{tab:3bodybg}
\end{table}

\begin{figure}
\resizebox{\textwidth}{!}{
\includegraphics*[0,0][550,550]{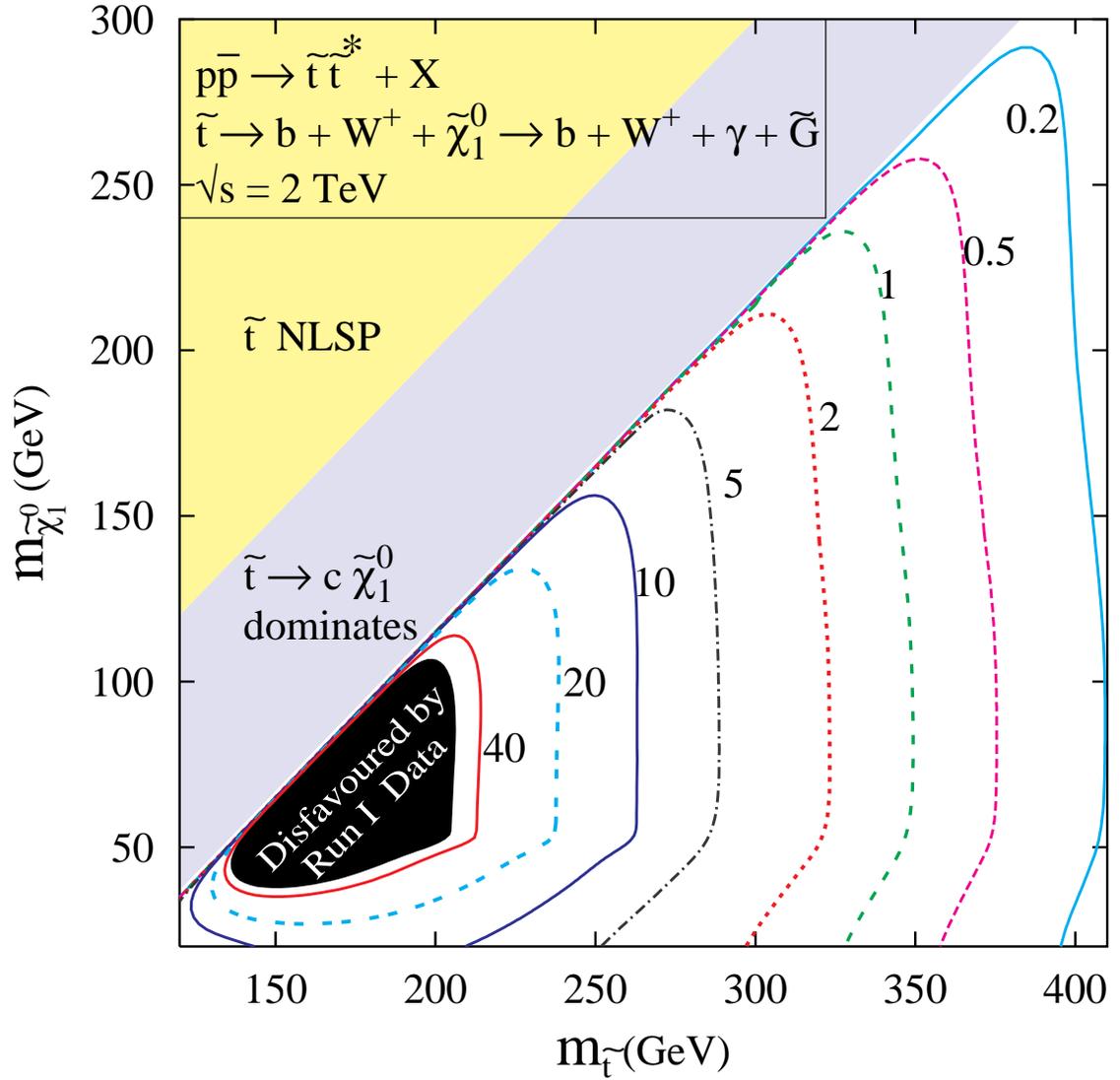}
}
\caption{Cross section in fb for stop pair production with 
$\tilde t \to b W \gamma \tilde G$, after cuts.
Both $W$ bosons are assumed to decay hadronically.
The black area is excluded by non-observation of $jjWW\gamma\gamma\missET$
events in Run I.
}
\label{fig:3bodyxsec}
\end{figure}

The signal cross section after cuts (but before
photon identification efficiencies) is shown in Fig.~\ref{fig:3bodyxsec}
as a function of the stop and $\tilde \chi_1^0$ masses.
We assume that the branching ratio of $\tilde t \to b W \tilde \chi_1^0$
dominates in the region of parameter space under consideration.  
The branching ratio of 
$\tilde \chi_1^0 \to \gamma \tilde G$ is again taken from Fig.~\ref{fig:binoBR}
assuming that $\tilde \chi_1^0$ is a pure bino.
The signal efficiency after cuts is about 45\%.
For small neutralino masses ($m_{\tilde \chi_1^0} \lsim 50$ GeV), 
though, both the photons and the gravitinos ($\missET$) tend
to be soft, leading to a decrease in the signal efficiency. 
For small stop masses
(as well as for large stop masses when the stop-neutralino 
mass difference is small), on the other hand, the jets are soft leading to 
a suppression of the signal cross section after cuts.  
As one would expect, both of these effects are particularly pronounced 
in the contours corresponding to large values of the cross section.
The additional distortion of the contours for large neutralino masses
can, once again, be traced to the suppression of the 
$\tilde \chi_1^0  \to \gamma \tilde G$ branching fraction 
(see Fig.~\ref{fig:binoBR}).

The non-observation of $jjWW\gamma\gamma\missET$ events at Run I of the 
Tevatron already excludes the region of parameter space shown in black
in Fig.~\ref{fig:3bodyxsec}.  As in the previous section, in this excluded
region at least 3 signal events would have been produced
after cuts and detector efficiencies in 
the 100 pb$^{-1}$ of Run I data, with negligible background.
In particular, Run I data excludes stop masses below about 200 GeV,
for neutralino masses larger than about 50 GeV.

We show in Table~\ref{tab:3bodyevents} the expected maximum stop
discovery mass reach at Tevatron Run II for various amounts of integrated
luminosity.\footnote{Again, 
if data from the CDF and D{\O} detectors are combined,
the integrated luminosity of the machine is effectively doubled.}
In particular, with 4 fb$^{-1}$, a stop discovery can
be expected in this channel if $m_{\tilde t} < 320$ 
GeV.\footnote{For 
comparison, in the case of minimal supergravity a reach of 
$m_{\tilde t} < 190$ GeV can be expected in the 
$\tilde t \to b W \tilde \chi_1^0$ channel with 4 fb$^{-1}$ at Tevatron
Run II \cite{Demina}.}
If the lightest neutralino is not a pure bino, the reach at large 
neutralino masses will be reduced.  However, the maximum stop mass reach
quoted here will not be affected, because it occurs for 
$m_{\tilde \chi_1^0} \sim 50 - 100$ GeV; in this mass range the lightest
neutralino decays virtually 100\% of the time to $\gamma \tilde G$ 
due to the kinematic suppression of all other possible decay modes, unless
its photino component is fine-tuned to be tiny.
\begin{table}
\begin{center}
\begin{tabular}{|c|c|c|c|c|}
\hline
$\int\mathcal{L}$ & B & S for a $5\sigma$ discovery & 
$\sigma_S \times \epsilon_{\gamma}^2$ 
   & Maximum stop mass reach \\
\hline
2 fb$^{-1}$  & 0.1 & 5  & 2.5 fb & 300 GeV\\
4 fb$^{-1}$  & 0.2 & 6  & 1.5 fb & 320 GeV\\
15 fb$^{-1}$ & 0.9 & 8  & 0.53 fb & 355 GeV\\
30 fb$^{-1}$ & 1.8 & 10 & 0.33 fb & 375 GeV\\
\hline
\end{tabular}
\end{center}
\caption{
Number of signal events (S) required for a $5\sigma$ stop discovery
at Tevatron Run II in the $jjWW\gamma\gamma \missET$ channel and the 
corresponding signal cross section after cuts and efficiencies and 
maximum stop mass reach.  
We take $\epsilon_{\gamma} = 0.80$.
The number of background events (B) is based on a background cross section
of 0.06 fb from Table~\ref{tab:3bodybg}.
}
\label{tab:3bodyevents}
\end{table}

Including a separate analysis of stop production and decay with 
one or more of the $W$ bosons decaying leptonically
would yield an increase in the overall signal statistics;
however, we do not expect this increase
to dramatically alter the stop discovery reach.

%-----------------------------------------------------------------------
\subsection{Stop production in top quark decays}
\label{sec:topdecay}

If $m_{\tilde t} + m_{\tilde \chi_1^0} < m_t$, then stops can be produced
in the decays of top quarks.  As we will explain here,
most of the parameter space in this region is excluded by the 
non-observation of stop events via direct production or in top quark 
decays at Run I of the Tevatron.  However, some interesting parameter 
space for this decay remains allowed after Run I, especially if the 
lighter stop is predominantly $\tilde t_L$.

For $m_{\tilde t} < m_b + m_W + m_{\tilde \chi_1^0}$, so that the stop
decays via $\tilde t \to c \tilde \chi_1^0$, the region in which
$t \to \tilde t \tilde \chi_1^0$ is possible is almost entirely 
excluded by the limit on stop pair production at Run I, as shown
in Fig.~\ref{fig:2bodyxsec}.
A sliver of parameter space in which the stop-neutralino mass splitting
is smaller than about 10 GeV remains unexcluded.
For $m_{\tilde t} > m_b + m_W + m_{\tilde \chi_1^0}$, so that the 
stop decays via $\tilde t \to b W \tilde \chi_1^0$,
the signal efficiency in the search for direct stop production 
is degraded for light neutralinos with masses below about 50 GeV
and for stops lighter than about 150 GeV.  This prevents Run I
from being sensitive to stop pair production in the region of 
parameter space in which top quark decays to stops are possible with
the stops decaying to $b W \tilde \chi_1^0$,
as shown in Fig.~\ref{fig:3bodyxsec}.
In what follows, we focus on this latter region of parameter space.

As discussed before, if the lightest neutralino is mostly bino, the 
constraints on its mass are model-dependent.  
The constraints from Tevatron Run I
are based on inclusive chargino and neutralino 
production~\cite{KaneTev,DZeroino}
under the assumption of gaugino mass unification;
the cross section is dominated by production of 
$\tilde \chi_1^{\pm} \tilde \chi_1^{\mp}$ and
$\tilde \chi_1^{\pm} \tilde \chi_2^0$.
If the assumption of gaugino mass unification is relaxed, 
then Run I puts no constraint on the mass of $\tilde \chi_1^0$.
At LEP, while the pair production of a pure bino leads to an easily
detectable diphoton signal, it proceeds only via $t$-channel 
selectron exchange.  The mass bound on a bino $\tilde \chi_1^0$ 
from LEP thus depends on the selectron mass \cite{LEPSUSYWG}.  
In particular, for selectrons heavier than about 600 GeV, 
bino masses down to 20 GeV are still allowed by the LEP data.
If $\tilde \chi_1^0$ contains a Higgsino admixture, it couples to the $Z$
and can be pair-produced at LEP via $Z$ exchange.
For an NLSP with mass between 20 and 45 GeV, as will be relevant
in our top quark decay analysis, the LEP search results 
limit the $\tilde H_2$ component to be less than 1\%.  
Such a small $\tilde H_2$ admixture has no appreciable effect
on the top quark partial width to $\tilde t \tilde \chi_1^0$.
We thus compute the partial width for $t \to \tilde t \tilde \chi_1^0$
assuming that the neutralino is a pure bino.
Taking the lighter stop to be 
$\tilde t_1 = \tilde t_L \cos\theta_{\tilde t} 
+ \tilde t_R \sin\theta_{\tilde t}$, we find,
\begin{equation}
        \Gamma(t \to \tilde t_1 \tilde B) 
        = \left[\frac{4}{9} \sin^2\theta_{\tilde t} 
        + \frac{1}{36}\cos^2\theta_{\tilde t} \right]
        \frac{\alpha}{\cos^2 \theta_W}
        \frac{E_{\tilde B}}{m_t} \sqrt{E_{\tilde B}^2 - m_{\tilde B}^2},
\label{eq:topwidth}
\end{equation}
where $E_{\tilde B} = (m_t^2 + m_{\tilde B}^2 - m_{\tilde t}^2)/2 m_t$.
The numerical factors in the square brackets in Eq.~\ref{eq:topwidth}
come from the
hypercharge quantum numbers of the two stop electroweak eigenstates.
Clearly, the partial width is maximized if the lighter stop is a
pure $\tilde t_R$ state; it drops by a factor of 16 if the lighter
stop is a pure $\tilde t_L$ state.
In any case, the branching ratio for $t \to \tilde t \tilde B$ 
does not exceed 6\% for $m_{\tilde t} > 100$ GeV and 
$m_{\tilde \chi_1^0} > 20$ GeV.\footnote{For neutralino masses below $m_Z$, 
as are relevant here, the branching ratio into $\gamma \tilde G$ 
is virtually 100\% (see Fig.~\ref{fig:binoBR}), almost independent
of the neutralino composition.}

\begin{figure}
\resizebox{\textwidth}{!}{\rotatebox{270}{\includegraphics{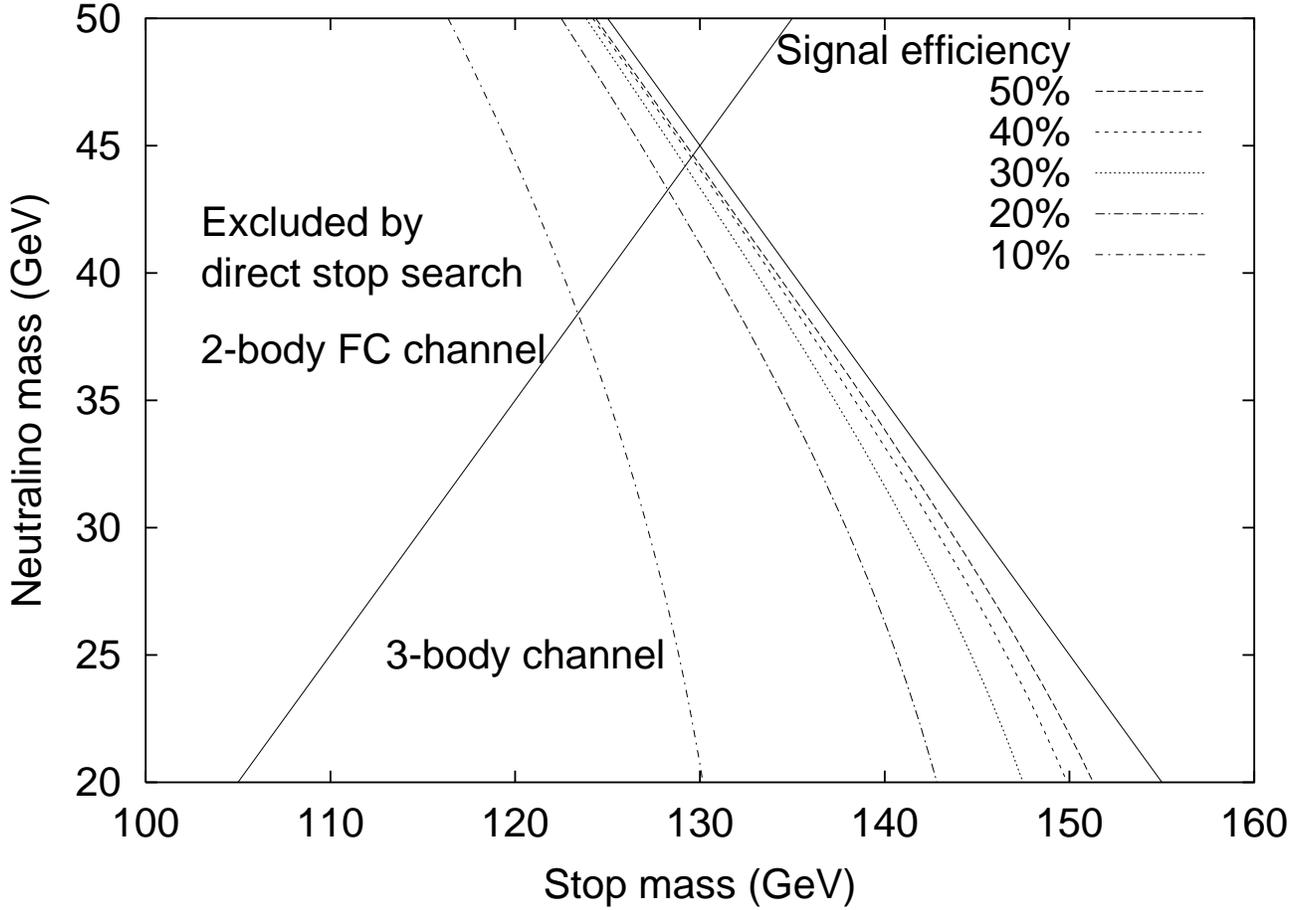}}}
\caption{95\% confidence level exclusion limits for top quark
decays to stops from Run I for various
signal efficiencies, in the case that $\tilde t_1 = \tilde t_R$, 
which gives the largest event rates.  
The area to the left of the curves is excluded.  
The case $\tilde t_1 = \tilde t_L$ gives no exclusion for the 
signal efficiencies considered and is not shown here.
The solid line from upper left to lower right 
is the kinematic limit for $m_t = 175$ GeV.  The solid line 
from upper right to lower left separates the regions in which the 2-body
FC decay and the 3-body decay of the stop dominate.}
\label{fig:topRun1}
\end{figure}

The signal from top quark pair production followed by one top
quark decaying as in the Standard Model and the other decaying to 
$\tilde t \tilde \chi_1^0$, followed by the stop 3-body decay and the
neutralino decays to $\gamma \tilde G$, is $bbWW\gamma\gamma\missET$.
This signal is the same (up to kinematics) as that from stop pair production 
in the 3-body decay region.
As in the case of stop pair production followed by the 3-body decay, 
we expect that the background to this process can be reduced to a 
negligible level.  
Run I can then place 95\% confidence level exclusion limits on 
the regions of parameter space in which 3 or more
signal events are expected after cuts and efficiencies are taken into account.
Using the Run I top quark pair production cross section of
6 pb \cite{topxsec} and a total luminosity of 100 pb$^{-1}$,
we compute the number of signal events as a function of the 
stop and neutralino masses and the stop composition, assuming various
values of the signal efficiency after cuts and detector efficiencies.
If $\tilde t_1 = \tilde t_R$, then Run I excludes most
of the parameter space below the kinematic limit for this decay
even for fairly low signal efficiency $\sim 20$\%,
as shown in Fig.~\ref{fig:topRun1}.
If $\tilde t_1 = \tilde t_L$, on the other hand, the signal cross section
is much smaller and Run I gives no exclusion unless the
signal efficiency is larger than 75\%, which would already be unfeasible
including only the identification efficiencies for the two photons;
even 100\% signal efficiency would only yield an exclusion up to 
$m_{\tilde t} \simeq 118$ GeV.

\begin{figure}
\resizebox{\textwidth}{!}{\rotatebox{270}{\includegraphics{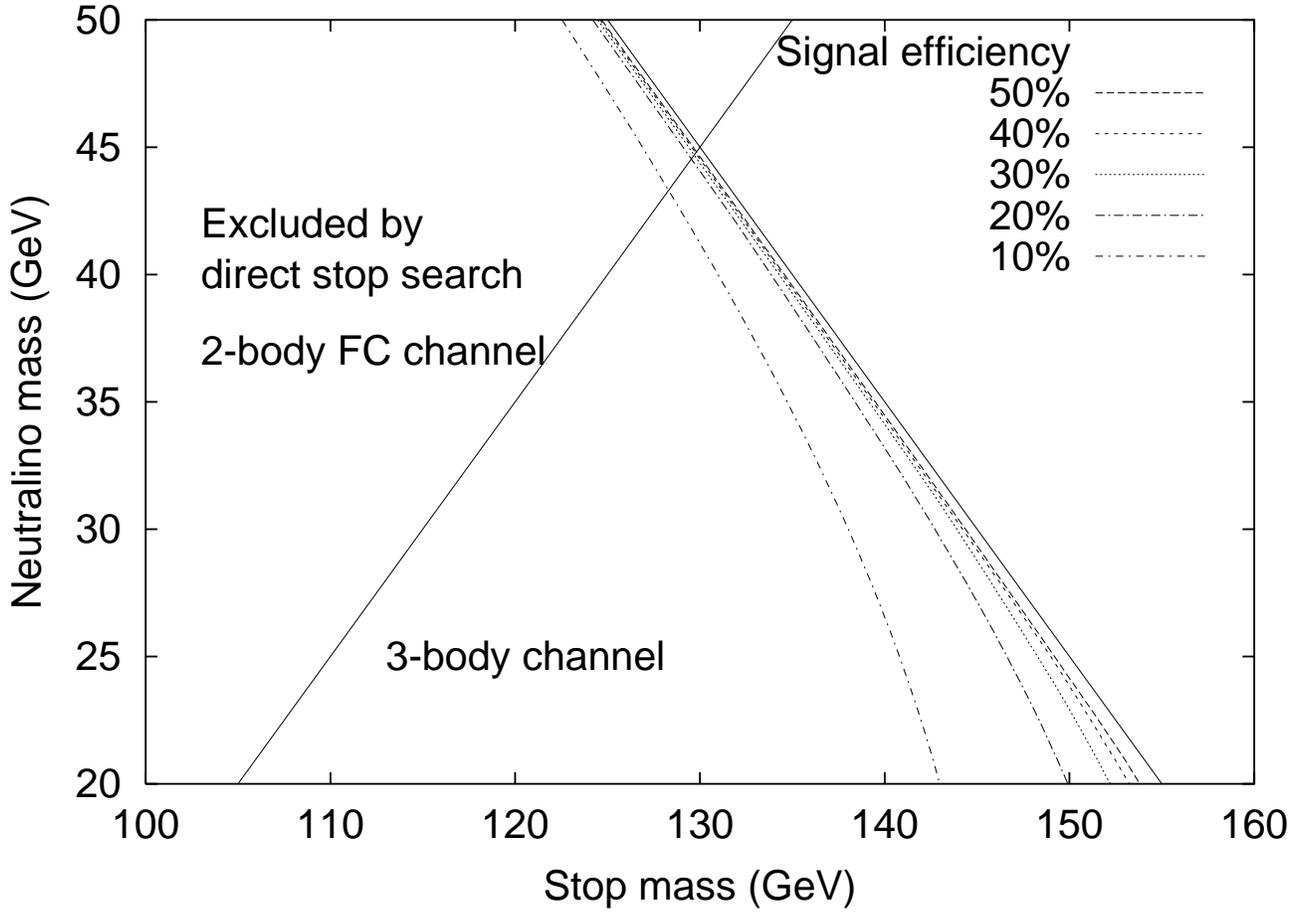}}}
\caption{$5\sigma$ discovery contours for top quark decays to stops
at Run II with 4 fb$^{-1}$ for various signal efficiencies, 
in the case that $\tilde t_1 = \tilde t_L$,
which gives the smallest event rates.
The case $\tilde t_1 = \tilde t_R$ would be discovered virtually
up to the kinematic limit even with 10\% signal efficiency, and
is not shown here.
The solid lines are as in Fig.~\ref{fig:topRun1}.}
\label{fig:topRun2}
\end{figure}

At Run II, the top quark pair production cross section is 
8 pb \cite{topxsec} and the expected total luminosity is considerably higher. 
This allows top quark decays to $\tilde t \tilde \chi_1^0$ 
to be detected for stop and neutralino masses above the Run I bound.
For $\tilde t_1 = \tilde t_R$, top quark decays to stops
will be probed virtually up to the kinematic limit,
even with low signal efficiency $\sim$10\%
and only 2 fb$^{-1}$ of integrated luminosity.  
If $\tilde t_1 = \tilde t_L$, so that the signal event rate is 
minimized, top quark decays to stops would be discovered 
up to within 10 GeV of the kinematic limit for signal efficiencies
$\gsim 20$\% and 4 fb$^{-1}$ of integrated luminosity 
(see Fig.~\ref{fig:topRun2}).
In this region of parameter space, stops would also be discovered in 
Run II with less than 2 fb$^{-1}$ via direct stop pair production
(see Fig.~\ref{fig:3bodyxsec}).

%-------------------------------------------------------------------
\section{Ten degenerate squarks}
\label{sec:10squarks}

Having concentrated until now on the production and decay of the top squark, 
let us now consider the other squarks. In the most general MSSM, the spectrum 
is of course quite arbitrary.  However, low energy constraints~\cite{fcnc} 
from flavor changing neutral current processes demand that such squarks 
be nearly mass degenerate, at least those of the same chirality. 
Interestingly, in many theoretical scenarios, such as 
the minimal gauge mediated models, this mass degeneracy between squarks
of the same chirality happens naturally; 
in addition the mass splitting between the 
left-handed and right-handed squarks associated with the
five light quarks turns out to be small. For simplicity, then, we will 
work under the approximation that all of these 10 squarks (namely,
$\tilde u_{L,R}$, $\tilde d_{L,R}$, $\tilde c_{L,R}$, $\tilde s_{L,R}$
and $\tilde b_{L,R}$) are exactly degenerate. 

\subsection{Production at Tevatron Run II}

While the cross sections for the individual pair-production of the 
$\tilde c_{L,R}$, $\tilde s_{L,R}$ and $\tilde b_{L,R}$ are 
essentially the same as that for a top-squark of the same mass, the 
situation is more complicated for squarks of the first generation. 
The latter depend sensitively on the gluino mass because of the presence
of $t$-channel diagrams. Moreover, processes such as 
$\bar u u \to \tilde u_L \tilde u_R^*$ or 
$d d \to \tilde d_{L, R} \tilde d_{L, R}$ become possible and relevant. 
Of course, in the limit of very large gluino mass, the squark production
processes are driven essentially by QCD and dominated  
by the production of pairs of mass eigenstates,
analogous to the top squark production considered
already.  In particular, at the leading order,
the total production cross section for the
ten degenerate squarks of a given mass is simply ten times the 
corresponding top squark production cross section.

\begin{figure}
\centerline{
\resizebox{\textwidth}{!}{
\rotatebox{270}{\includegraphics{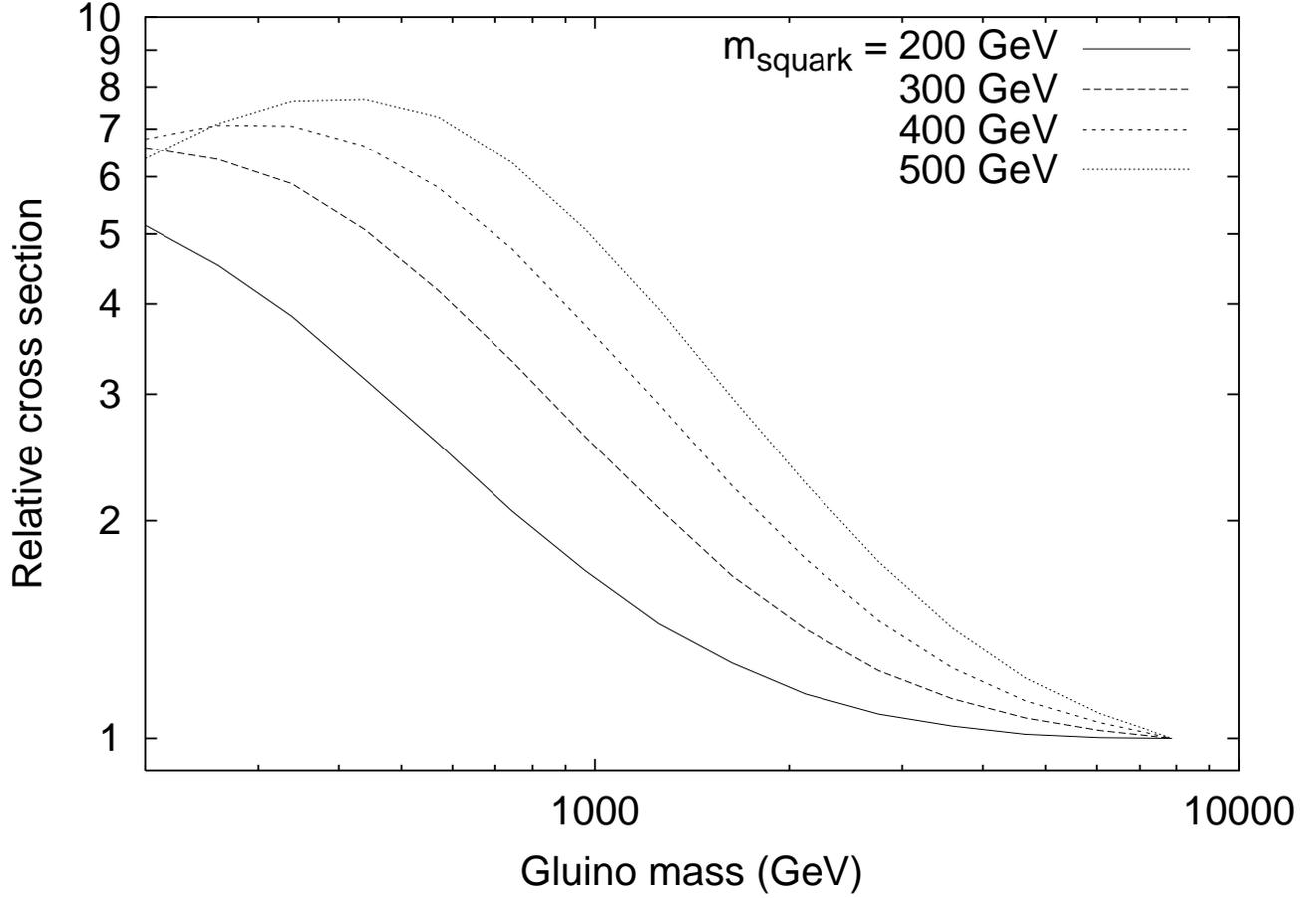}}
}}
\caption{Gluino mass dependence
of the NLO cross section for production of ten degenerate squarks
in 2 TeV $p \bar p$ collisions, from PROSPINO~\cite{PROSPINO}.
Shown are the cross sections for $\tilde q \tilde q^*$ production
normalized to the value at large $m_{\tilde g}$, for common
squark masses of 200, 300, 400 and 500 GeV.
Production of $\tilde q \tilde q$ is small at a $p \bar p$ collider
and is neglected here; it yields an additional 3-15\% increase 
in the total cross section at low $m_{\tilde g}$ for this range 
of squark masses.
Cross sections are evaluated at the scale $\mu = m_{\tilde q}$.
}
\label{fig:10sqgluino}
\end{figure}

Since a relatively light gluino only serves to increase the total 
cross section (see Fig.~\ref{fig:10sqgluino}), it can be argued that the 
heavy gluino limit is a {\em conservative one}. To avoid considering an 
additional free parameter, we shall perform our analysis in this limit. 
To a first approximation, the signal cross sections presented below will 
scale\footnote{That the gluino exchange diagram has a different 
        topology as compared to the (dominant) quark-initiated QCD diagram 
        indicates that corresponding angular distributions would be somewhat 
        different. Thus, the efficiency after cuts is not expected to 
        be strictly independent of the gluino mass. For the most 
        part, though, this is only a subleading effect.}
with the gluino mass approximately as shown in 
Fig.~\ref{fig:10sqgluino}.

Like top squarks, the 10 degenerate squarks can also be produced via 
cascade decays of heavier supersymmetric particles.  To be 
conservative, we again neglect this source of squark production
by assuming that the masses of the heavier supersymmetric particles
are large enough that their production rate at Tevatron energies can
be neglected.

\begin{figure}
\resizebox{\textwidth}{!}{
\rotatebox{270}{\includegraphics{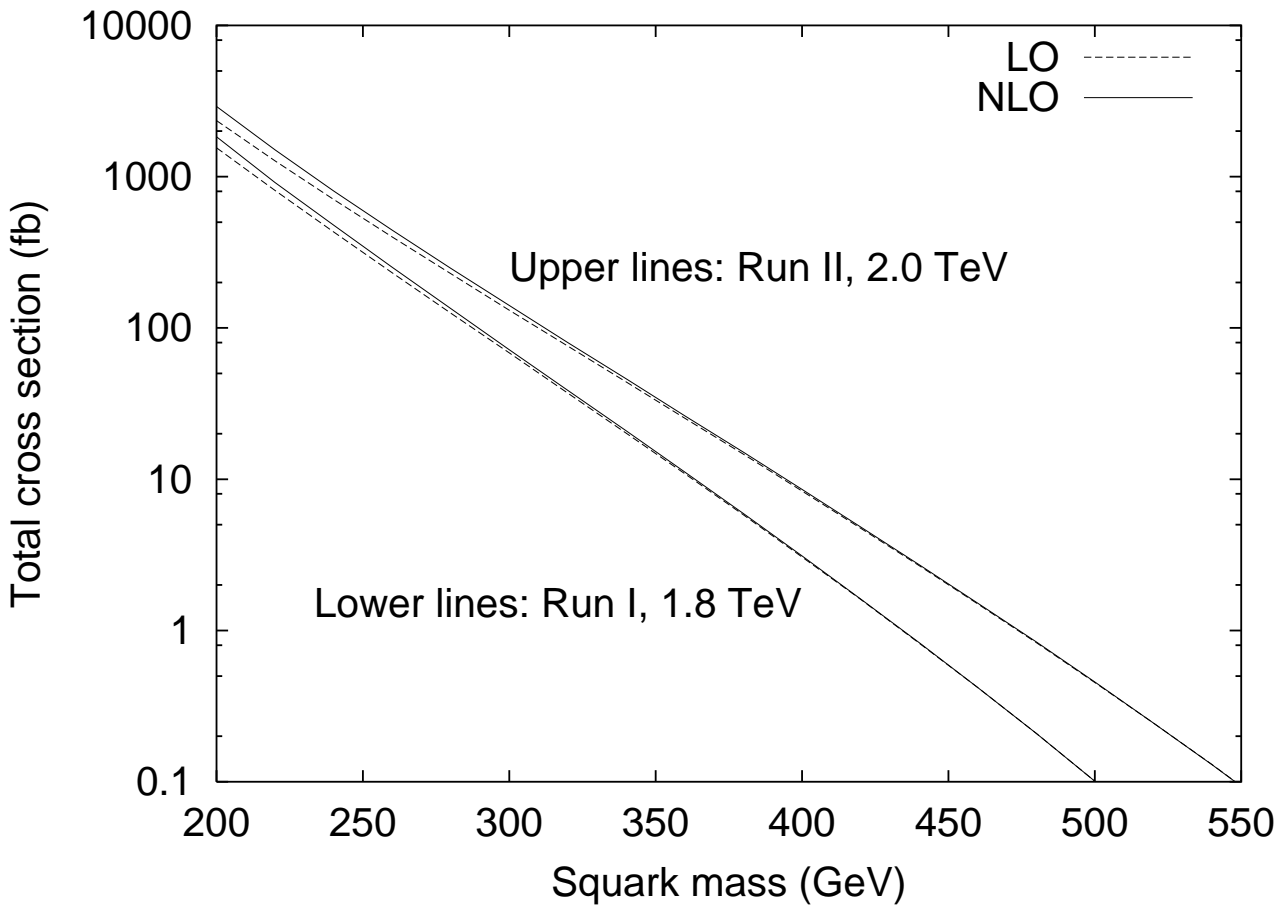}}
}
\caption{LO and NLO cross section 
for the production of ten degenerate squarks in the heavy gluino limit
in $p \bar p$ collisions at Tevatron Run I (1.8 TeV) and Run II (2.0 TeV), 
from PROSPINO~\cite{PROSPINO}.
Cross sections are evaluated at the scale $\mu = m_{\tilde q}$.
}
\label{fig:10sqxsecmass}
\end{figure}

The NLO cross sections for production of ten degenerate squarks 
including QCD and SUSY-QCD corrections
have been implemented numerically in PROSPINO~\cite{PROSPINO}.
We generate squark production events using the LO cross section evaluated
at the scale $\mu = m_{\tilde q}$, improved by the NLO K-factor obtained
from PROSPINO~\cite{PROSPINO} (see Fig.~\ref{fig:10sqxsecmass}), 
in the limit that the gluino is very heavy.  
The K-factor varies between 1 and 1.25 for $m_{\tilde q}$
decreasing from 550 to 200 GeV.
As in the case of the top squark analysis, we
use the CTEQ5 parton distribution functions~\cite{CTEQ5} and neglect 
the shift in the $p_T$ distribution of the squarks due to gluon radiation
at NLO.

%-----------------------------------------------
\subsection{Signals in low-energy SUSY breaking}

The decays of the ten degenerate squarks are very simple. 
As long as they are heavier than the lightest neutralino,\footnote{As 
        before, we do not allow the possibility
        of cascade decays through other neutralinos/charginos.
        Were we to allow these, the channel we are considering would 
        be somewhat suppressed, but additional, more spectacular,
        channels would open up.}
they decay via $\tilde q \to q \tilde \chi_1^0 \to q \gamma \tilde G$.  
The signal and
backgrounds are then identical to those of the two-body FC stop decay 
discussed in Sec.~\ref{sec:2body}, and consequently we use the same
selection cuts.
In fact, in view of the tenfold increase in the 
signal strength, we could afford more stringent cuts so as to 
eliminate virtually all backgrounds, but this is not quite necessary.

\begin{figure}
\resizebox{\textwidth}{!}{
\includegraphics*[0,0][550,550]{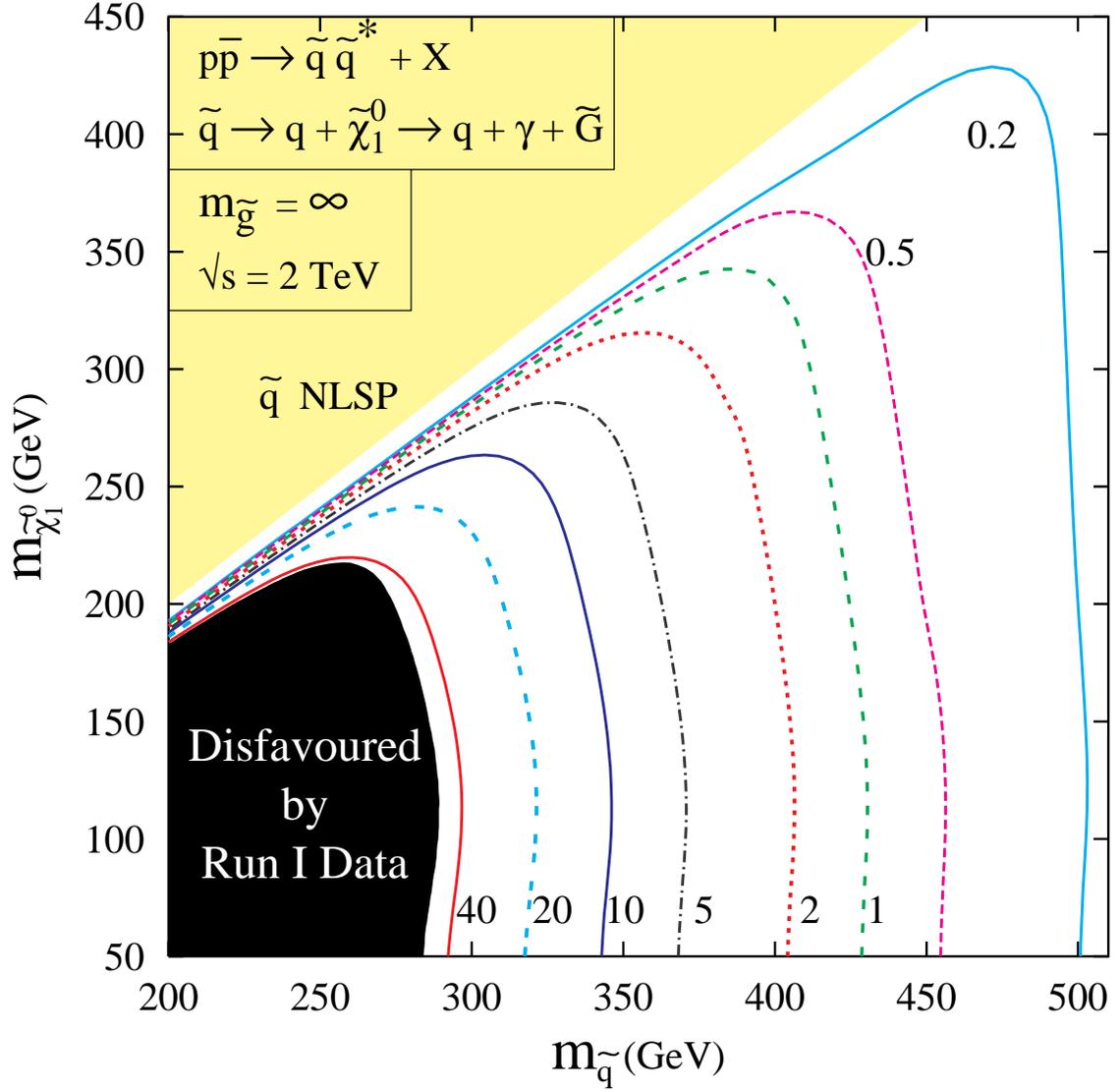}
}
\caption{Cross section in fb for production of 10 degenerate squarks
in Run II in the heavy gluino limit with 
$\tilde q \to q \gamma \tilde G$, after cuts.
The black area is excluded by non-observation of $jj\gamma\gamma\missET$
events in Run I.
}
\label{fig:10squarks}
\end{figure}

The signal cross section after cuts (but before efficiencies)
for production of ten degenerate squarks
in the heavy gluino limit is shown in Fig.~\ref{fig:10squarks}
as contours in the $m_{\tilde q}$--$m_{\tilde \chi_1^0}$ plane.
We assume that the squark $\tilde q$ decays predominantly into 
$q \tilde \chi_1^0$.  The branching ratio of 
$\tilde \chi_1^0 \to \gamma \tilde G$
is taken from Fig.~\ref{fig:binoBR} assuming that $\tilde \chi_1^0$ is 
a pure bino.
Clearly, the effect of the kinematic cuts on the signal is 
very similar to that in the case of the 2-body FC decay of the stop. 
The mass reach, of course, is much larger due to the tenfold increase in 
the total cross section; also, unlike in the case of the stop, 
the 2-body decay is dominant throughout the entire parameter space.
        
The non-observation of $jj\gamma\gamma \missET$ events at Run I of the
Tevatron excludes the region of parameter space shown in black in 
Fig.~\ref{fig:10squarks}.  
As in our stop analysis, in this excluded
region, at least 3 signal events would have been detected in 
the 100 pb$^{-1}$ of Run I data, with negligible background.  
In particular, we estimate that Run I data excludes the ten degenerate 
squarks up to a common mass of about 280 GeV.  
As in the case of 2-body FC stop decays, when the 
mass splitting between the squarks and the neutralino is too small 
({\it i.e.}, less than about 40 GeV), 
the jets become soft and the signal efficiency decreases dramatically,
leaving an unexcluded region of parameter space.
The D{\O} search for inclusive $p \bar p \to \tilde \chi_2^0 + X$ 
with $\tilde \chi_2^0 \to \gamma \tilde \chi_1^0$ \cite{DZerosquark}
yields a limit on the production cross section of about 0.5 pb for
parent squark masses above about 250 GeV.
Interpreted in terms of pair production of 10 degenerate squarks in 
the large $m_{\tilde g}$ limit with 
$\tilde \chi_2^0 \to \gamma \tilde \chi_1^0$ reidentified as
$\tilde \chi_1^0 \to \gamma \tilde G$ again increases the signal efficiency
by a factor of $\sim 2.7$ because every event contains two photons 
\cite{DZerosquark}; the D{\O} analysis then yields a bound on the 
common squark mass of about 275 GeV, again in rough agreement with
our result.\footnote{Ref.~\cite{DZerosquark} quotes a squark mass bound of 
320 GeV for the case $m_{\tilde q} \ll m_{\tilde g}$
in the context of low-energy SUSY breaking; we expect that this is due
to the contribution of chargino and neutralino production to the 
total SUSY cross section in their analysis.}

Taking the total
background cross section to be 2 fb (see Table~\ref{tab:2bodybg}), 
we show in Table~\ref{tab:10squarkevents} the expected maximum
discovery mass reach for ten degenerate squarks at Tevatron Run~II 
for various amounts of integrated
luminosity.\footnote{Again, 
if data from the CDF and D{\O} detectors are combined,
the integrated luminosity of the machine is effectively doubled.}
In particular, with 4 fb$^{-1}$ a squark discovery can
be expected in this channel if $m_{\tilde q} < 360$ GeV.
Again, if the lightest neutralino is not a pure bino, the reach 
at large neutralino masses will be reduced.  However, this will
have very little effect on the maximum squark mass reach quoted here,
because this maximum reach occurs for $m_{\tilde \chi_1^0} \sim 100 - 150$ GeV,
where all neutralino decay modes other than $\gamma \tilde G$ suffer
a large kinematic suppression.  From Fig.~\ref{fig:binoBR}, it is 
evident that the neutralino branching fraction into $\gamma \tilde G$
will be reduced by no more than 10\% in this mass range as long as
the neutralino is at least 50\% bino; such a reduction in the neutralino
branching fraction will lead to a reduction of only a few GeV in the maximum 
squark mass reach.

\begin{table}
\begin{center}
\begin{tabular}{|c|c|c|c|c|}
\hline
$\int\mathcal{L}$ & B & S for a 5$\sigma$ discovery 
   & $\sigma_S \times \epsilon_{\gamma}^2$
   & Maximum squark mass reach \\
\hline
2 fb$^{-1}$  &  4 & 14 &  7.0  fb & 345 GeV\\
4 fb$^{-1}$  &  8 & 18 &  4.5 fb & 360 GeV\\
15 fb$^{-1}$ & 30 & 31 &  2.1 fb & 390 GeV\\
30 fb$^{-1}$ & 60 & 42 &  1.4 fb & 405 GeV\\
\hline
\end{tabular}
\end{center}
\caption{
Number of signal events (S) required for a 5$\sigma$ squark discovery at 
Tevatron Run~II, assuming production of 10 degenerate squarks in the limit
that the gluino is very heavy,
and the corresponding signal cross section after cuts and efficiencies 
and maximum squark mass reach.  We take $\epsilon_{\gamma} = 0.80$.
The number of background events (B) is based on a background cross section 
of 2 fb from Table~\ref{tab:2bodybg}.
}
\label{tab:10squarkevents}
\end{table}

%------------------------------------------------------------------
\section{Conclusions}
\label{sec:conclusions}

In models of low-energy SUSY breaking, signatures of SUSY particle production
generically contain two hard photons plus missing energy due to the decays
of the two neutralino NLSPs produced in the decay chains.  Standard Model
backgrounds to such signals are naturally small at Run II of the Tevatron.
We studied the production and decay of top squarks at the Tevatron in such
models in the case where the lightest Standard Model superpartner is
a light neutralino that predominantly decays into a photon and a 
light gravitino.
We considered 2-body flavor-changing and 3-body decays of the top squarks.
The reach of the Tevatron
in such models is larger than in the standard supergravity models
and than in models with low-energy SUSY breaking in which the stop is
the NLSP, rather than the neutralino.
We estimate that top squarks with masses below about 200 GeV
can be excluded based on Run~I data, assuming that 
50 GeV $\lsim m_{\tilde \chi_1^0} \lsim m_{\tilde t} - 10$ GeV. 
For a modest final Run II luminosity of 4 fb$^{-1}$, stop masses 
up to 285 GeV are accessible in the 2-body decay mode, and up 
to 320 GeV in the 3-body decay mode.

Top squarks can also be produced in top quark decays.  We found that,
within the context of low-energy SUSY breaking with the stop as the
next-to-next-to-lightest SUSY particle,
the region of parameter space in which stop production in top quark decays 
is possible is almost entirely excluded by Run I data
if the lighter stop is predominantly right handed; however, an interesting
region is still allowed if the lighter stop is predominantly left handed,
due to the smaller branching ratio of $t \to \tilde t_L \tilde \chi_1^0$.
Run II will cover the entire parameter space in which top decays to stop
are possible.

We also studied the production and decay of the ten squarks associated
with the five light quarks, assumed to be degenerate.
In models of low-energy SUSY breaking, the decays of the ten degenerate
squarks lead to signals identical to those for the 2-body flavor-changing 
stop decays.
The cross section for production of ten degenerate squarks at the Tevatron
is significantly larger than that of the top squark.
We estimate that the 10 degenerate squarks with masses below about
280 GeV can be excluded based on Run~I data, assuming that
$m_{\tilde \chi_1^0} \lsim m_{\tilde q} - 40$ GeV.
For a final Run II luminosity of 4 fb$^{-1}$, squark masses as large as
360 GeV are easily accessible in the limit that the gluino is very heavy.  
The production cross section, and hence the discovery reach, 
increases further with decreasing gluino mass.

\vskip1cm
%---------------------------
\noindent
{\Large \bf Acknowledgments}

\vskip0.5cm \noindent
We are very grateful to Michael Schmitt for helping us understand
instrumental backgrounds and detector capabilities.
We also thank Ray Culbertson, Joel Goldstein, Tilman Plehn, 
and David Rainwater for useful discussions.
Fermilab is operated by Universities Research Association Inc.\
under contract no.~DE-AC02-76CH03000 with the U.S. Department of
Energy.  D.C. thanks the Theory Division of Fermilab for hospitality while 
part of the project was being carried out and 
the Deptt. of Science and 
Technology, India for financial assistance under the 
Swarnajayanti Fellowship grant.
R.A.D.\ thanks Fermilab for its kind hospitality and 
financial support and DINAIN (Colombia) and
COLCIENCIAS (Colombia) for financial support.
C.W. is supported in part by the US DOE, Division of High-Energy Physics,
under contract no.~W-31-109-ENG-38.

%\clearpage

%-------------------------------------------------------

\end{document}